\documentclass{ptephy_v1}

\usepackage{bm}
\usepackage{graphics} 
\usepackage{subfig} 

\begin{document}

\title{Saturation of Nuclear Matter and Roles of Many-Body Forces: 
      -- nuclear matter in neutron stars probed by nucleus-nucleus scattering --}


\author{Yukinori Sakuragi\thanks{The given name is Hiroyuki Sakuragi}}
\affil{Department of Physics, Osaka City University, Osaka 558-8585, Japan \email{sakuragi@sci.osaka-cu.ac.jp}}


\begin{abstract}%
Yoichiro Nambu put a great foot print in nuclear physics in the era of its fundamental developments including his pioneering insight into essential ingredients of repulsive core of nuclear force and its relation to the saturation of nuclear matter. 
The present review article focuses onto recent developments of the interaction models between colliding nuclei in terms of Brueckner's $G$-matrix theory starting from realistic nuclear forces and the saturation property of symmetric nuclear matter as well as neutron-star matter. 
A recently proposed unique scenario of extracting the saturation property of nuclear matter and stiffness of neutron stars through the analysis of nucleus-nucleus elastic scattering in laboratories is presented in some detail.
\end{abstract}

\subjectindex{xxxx, xxx}

\maketitle

\section{Introduction}

Yoichiro Nambu put great foot prints in physics.
The greatest achievement will be his genius idea of spontaneous symmetry breaking (SSB) inspired by the BCS theory for the superconductivity in condensed matter.
The famous Nambu-Jona-Lasinio (NJL) papers~\cite{SBB1,SBB2} marked a great milestone in particle and nuclear physics.
The idea proposed in the NJL papers has lead to the idea of dynamical breaking of chiral symmetry in Quantum Chromodynamics (QCD) of the standard model and the original NJL model is now understood as a low-energy effective theory of QCD of highly non-perturbative nature. 

Since Nambu's contributions to various fields in physics are reviewed by other authors in the review articles for the present memorial symposium, this paper will be devoted to some specific aspects of his contribution, particularly to the repulsive nature of nuclear force and the saturation property of nuclear matter, and its related topics relevant to the saturation property of nuclear matter probed by the nucleus-nucleus scattering will be presented.

To be more specific, we introduce a recently proposed unique scenario of extracting  important information about the saturation property of nuclear matter and the stiffness of neutron stars through the analysis of nucleus-nucleus elastic scattering in laboratories.
Key ingredients of the scenario are the short-range repulsive three-body forces that acts among nucleons and baryons in nuclear and neutron-star matter. 
The idea is that the unknown strength of the three-body force can be fixed quite precisely through the analyses of elastic scattering experiments in laboratories by the latest microscopic theory for interactions between finite nuclear systems. 

The neutron stars J1614-2230~\cite{NS10} and J0348-0432~\cite{NS13}
have brought great impacts on the maximum-mass problem,
observed masses of which are (1.97 $\pm$ 0.04)$M_\odot$ 
and (2.01 $\pm$ 0.04)$M_\odot$, respectively, much larger than what was observed before, say, (1.4 $\sim$ 1.5)$M_\odot$.
These large masses have added a severe condition for the stiffness of the equation of state (EOS) of neutron-star matter, that serves a great challenge to nuclear physics. 

In order to clarify the implication of the scenario, we briefly review a microscopic theory of interaction between finite nuclei, called the double-folding model (DFM) and the Brueckner's $G$-matrix theory, that are the keys to connect the infinite nuclear matter, including the baryonic matter in neutron stars, and the nucleus-nucleus scattering experiments in laboratories. 
To this end, we first present a brief history of the theoretical study of effective nucleon-nucleon (NN) interaction in nuclear medium derived from the Brueckner's $G$-matrix theory and its application in constructing the nucleus-nucleus interactions in terms of various types of the DFMs.

\section{Basic property of finite nuclei}
Before discussing the latest status of the effective interactions and its application to the nucleus-nucleus scattering and the saturation of nuclear matter, we first depict the basic property of atomic nuclei.

\subsection{Nuclear force and repulsive core}

The origin of the ``nuclear force'' (or the nucleon-nucleon (NN) interaction) that sustain protons and neutrons in forming an atomic nucleus having an extremely small size of radius less than 10$^{-14}$ m = 10 fm was first proposed by Hideki Yukawa~\cite{Yukawa35} as the exchange of a pion ($\pi$ meson) between nucleons (the one-pion exchange OPE).
The OPE is known to generate the most outside (long-range) part of the nuclear force because of the light mass of the exchanged pion, $m_\pi \simeq 140$ MeV, with the so-called Yukawa tail $\sim e^{-m_\pi r}/r$ of the nuclear force. 
The nuclear force has a more strong attraction in the middle range  ($0.5 \sim 1.0$ fm) region that is now understood by the exchange of heavier mesons such as $\sigma$, $\omega$ and $\rho$ mesons as well as by the exchange of multi pions.

In addition to the attraction in the middle and long-range parts of the nuclear force, a series of experimental measurements of nucleon-nucleon scattering at high energies indicate the existence of a strong repulsion at short distances less than about 0.5 fm.
The simplest version of the repulsive part of nuclear force is the so-called {\em hard core} of an infinite strength, that was first introduced phenomenologically by Jastrow in 1950~\cite{Jastrow51} to explain the nucleon-nucleon scattering experiments at high energies. To date, various types of the repulsive core, either of the hard-core type or the soft-core one, have been implemented in realistic nuclear force models widely used in nuclear structure and reaction studies, although the origin of the short-range repulsive core has long been unsettled.

One of the pioneering work for a possible origin of the short-range core of the nuclear force was made by Nambu who proposed a vector meson called the omega ($\omega$) meson in 1957~\cite{Nambu_Omega} as a possible source of the repulsive core, in addition to the source of the attraction in the middle range of the nuclear force.
The pioneering work by Nambu on the origin of the repulsive core was followed by various types of quark models~\cite{NST77,OY80} but it is only recently that a milestone of understanding the origin of the nuclear force, including the repulsive core, is marked in the epoch-making work by Ishii and his collaborators~\cite{Ishii07} with the first-principle calculations based on QCD, the standard model for strong interaction. 
They derived the radial shape and strength of nuclear force in the lattice QCD calculations that well reproduce the basic properties of nuclear force, not only in the qualitative level but also in quantitative level,  derived from the meson-exchange mechanism supplemented by a phenomenological repulsive core at short distances.

\subsection{Nuclear force and saturation of nuclear matter}

The nuclear radius observed in various experiments is found to be proportional to the cubic root of the mass number $A ( =Z+N)$, $R \simeq r_0 \times A^{1/3}$, 
which implies that nuclear volume $V=4\pi R^3/3$ is proportional to $A$ and, hence, the average  nucleon density in the nucleus $\rho=A/V=3/(4\pi r_0^3)$ is approximately constant, irrespective of the mass number $A$ of nuclei. 
The constant value $\rho_0 \simeq 0.17$ fm$^{-3}$ is called as the {\em saturation density}.
The mass of an atomic nucleus, $M(Z,N)$, with an atomic number $Z$ and a neutron number $N$ defines the binding energy of the nucleus 
by $B(Z,N)=\bigm( Z m_p+N m_n - M(Z,N) \bigm) c^2$, $m_p$ and $m_n$ being the mass of a proton and a neutron respectively. 
The experimental measurement of nuclear masses shows that the binding energy per nucleon 
$\bar{E}=B(Z,N)/A$ has an almost constant value of $\bar{E}\simeq 8 \pm 0.5$ MeV irrespective of the number of nucleons in nuclei, that is called the 
{\it saturation of binding energy} in atomic nuclei.
This indicates that the total binding energy of a nucleus is approximately proportional to its mass number $A$ (and hence to its volume $V$) as $B(Z,N)\approx a_V\times A$ MeV.

The saturation of the nucleon density and that of the binding energy per nucleon in atomic nuclei are closely related to the nature of nuclear force, particularly of its short-range nature and the existence of the short-range repulsive core.
From the following intuitive consideration, it is easily understood that saturation of the binding energy per nucleon is a direct consequence of the saturation of nucleon density in nuclei and the short-ranged nature of the nuclear force.

First of all, one should note that the maximum range of the nuclear force of about 
$1.5 \sim 2.0$ fm  is approximately the same order of the mean distance between nucleons in atomic nuclei evaluated from the saturation density of $\rho_0 \simeq 0.17$ fm$^{-3}$.
Because of the constant nucleon density in a nucleus (except in the region near nuclear surface), the number of nucleons surrounding a nucleon in the uniform nuclear matter is also constant.
This indicates that the binding energy for a nucleon by its surrounding nucleons in the uniform nuclear matter must be almost constant because the nuclear force is of short-range nature and does not act beyond the neighboring nucleons.
In other words, the total binding energy of nucleons in a nucleus should be proportional to the number of nucleons (mass number $A$) that is also proportional to the nuclear volume because $V \propto R^3 \propto A$. 
This term of the binding energy is then called the volume term. 

One might expect that the proportional constant $a_V$ of the volume term would be identified with the measured constant value of the binding energy per nucleon, 
$a_V \approx \bar{E}\simeq 8 \pm 0.5$ MeV.
However, this is not true at all and the measured constant value $8 \pm 0.5$ MeV is known to be a consequence of large corrections due to the finiteness of nuclei (or the effect of surface tension),  the Coulomb energy and the symmetry energy originating from the difference of neutron and proton number in a nucleus, all of which are the correction for reducing the binding energy.
Subtracting those correction terms, the genuine volume term of the binding energy per nucleon is found to be $a_V\simeq 16$ MeV in the symmetric nuclear matter of infinite extension in space, namely having no surface.
The binding energy per nucleon  $\bar{E} \simeq 16$ MeV is called the {\em saturation energy} of  the symmetric nuclear matter.

The saturation of the nucleon density and that of the binding energy per nucleon in atomic nuclei suggest the existence of a fundamental origin of sustaining the atomic nuclei in their size and binding and of preventing the nuclei from shrinkage or collapse due to the strong attraction of nuclear force. 
It is the short-range repulsion of the nuclear force that is one of the candidates of the origin of the saturation mentioned above.
However, the short-range repulsion itself of the nuclear force between two interacting nucleons, namely the two-body repulsion, is not sufficient for explaining the saturation property observed in nuclear experiments as well as  explaining the maximum mass of neutron stars $M\cong 2M_\odot$ obtained from the latest astronomical observations~\cite{NS10,NS13}.
It will be discussed as a major subject of the present paper that the effect of many-body forces of repulsive nature plays a decisive role to realize the saturation of nuclear matter as well as of the neutron-star matter.

\section{Effective interaction in nuclear medium}

Understanding the nuclear structure and reactions microscopically starting from the basic nucleon-nucleon (NN) interaction has been the fundamental and major subject in nuclear physics.
The basic properties of NN interaction in free space, say its shape, strength and their spin-isospin dependence, are well understood phenomenologically from the observed properties of a deuteron, that is the only bound state of the two nucleon system, and from NN scattering experiments at wide-range of scattering energies with various spin orientations of colliding nucleons. 
A number of theoretical models for the NN interaction in free space were also proposed, ranging from purely phenomenological models to those based on various meson theories and quark models.
A recent derivation of the nuclear force in terms of the first-principle lattice QCD calculation~\cite{Ishii07} based on the standard model for the strong interaction QCD marks a milestone of understanding the nuclear force, as mentioned in Introduction.

However, it should be emphasized that the interaction between nucleons in nuclei is not equivalent to that in free space and is strongly modified by the surrounding nucleons. 
This is called the {\em medium effect} or, in other words, the {\em many-body effect}.
The NN interaction modified in the nuclear medium is called the {\em effective NN interaction}. 

\subsection{Brueckner's $G$-matrix and saturation properties of nuclear matter}

The Brueckner's $G$-matrix theory~\cite{Brue55,Gmat95} is one of the most successful theories for handling the short-range correlation (SRC) of the bare NN interaction and generating an effective NN interaction in nuclear medium.
In the $G$-matrix theory, a pair of nucleons are embedded in infinite nuclear matter with a uniform nucleon density and one solves the ``two-body'' Schr\"{o}dinger equation,  called the {\em Bethe-Goldstone equation}, 
\begin{equation}
\bigm( \hat T_1 + \hat T_2 + U_1 + U_1 -E \bigm) \psi = \hat Q\: v_{12}\: \psi
\end{equation}
with the Pauli-blocking operator $\hat Q$ that prevents the interacting two nucleons from sitting in the states already occupied by other nucleons of the surrounding nuclear matter.
In other words, the interacting nucleon pair are virtually excited to unoccupied highly exited states in the intermediate states in the series of the ladder diagrams under the effect of Pauli blocking by the occupied nucleons in nuclear matter.
This implies that the motion of the pair nucleons are strongly restricted, though indirectly, by the surrounding nucleons through the Pauli-blocking operator $Q$, paticularly at high density nuclear matter.
In this sense, the medium effect is interpreted as the many-body effect.

Let us consider a pair of nucleons in momentum space, one being a bound nucleon with momentum $\bm{p}$ (in the unit of $\hbar$) in nuclear matter with the Fermi momentum $k_F$ and the other being an incoming nucleon with momentum $\bm{k}$, where  $|\bm{p}| < k_F$ and $|\bm{k}| > k_F$ in the initial states before the interaction. 
It is noted that the Fermi momentum $k_F$ is proportional to the qubic square of the 
uniform density $\rho$ of the nuclear matter, $k_F \propto \rho^{1/3}$.
Relative and center-of-mass momenta are given as $\bm{k}_{r}=(\bm{k}-\bm{p})/2$ and $\bm{K}_c=\bm{k}+\bm{p}$, respectively. With a NN interaction $V$, the $G$-matrix equation giving the scattering of the pair in the medium is represented as 
\begin{eqnarray}
G(\omega)=V+\sum_{\bm{q}_1,\bm{q}_2}V\frac{Q(\bm{q}_1,\bm{q}_2)}
{\omega-e(q_1)-e(q_2)+i\varepsilon}G(\omega) \ ,
                \label{eq:gmat0}
\end{eqnarray}
where $e(q)$ is a single particle (s.p.) energy in an intermediate state with momentum $q$, and $Q(\bm{q}_1,\bm{q}_2)$ is the above-mentioned Pauli blocking operator defined by 
\begin{eqnarray}
Q(\bm{q}_1,\bm{q}_2)\, |\bm{q}_1,\bm{q}_2 \rangle=
\left\{ \begin{array}{cl} |\bm{q}_1,\bm{q}_2 \rangle 
& \mbox{if $q_1,q_2>k_F$} \ , \\
0 & \mbox{otherwise} \ .
\end{array}
\right. 
\end{eqnarray}
The starting energy $\omega$ is given as a sum of the energy $E(k)$ of the propagating nucleon and the single-particle (s.p.) energy $e(p)=\frac{\hbar^2}{2m}p^2+U(p,e(p))$ of a bound nucleon~\cite{FUR08}. 

The $G$-matrix calculations are often performed with the so-called continuous choice for intermediate nucleon spectra, playing an essential role especially for imaginary parts of $G$-matrices. 
This choice means that s.p. energies 
\begin{eqnarray}
e(q)=\frac{\hbar^2}{2m}q^2+U(q,e(q))
\end{eqnarray}
are calculated self-consistently not only for bound states ($q \le k_F$) but also for intermediate states ($q>k_F$). 
The scattering boundary condition $i\varepsilon$ in the dominator leads to complex $G$-matrices, summation of which gives the complex s.p. potential $U(q,e(q))$. 
The plausible way is to use this self-consistent complex potential in $G$-matrix equation (\ref{eq:gmat0}). 
Avoiding numerical complexities in such a procedure, however, its real part $U_R(q)={\rm Re}\,U(q,e(q))$ is usually used in the self-consistency process~\cite{BR77a,BR77b,BR78,CEG83,CEG86}. 

The $G$-matrix also depends on the NN interaction model in free space $V$ adopted in the calculation, because of the difference in the properties of central, spin-orbit and tensor forces and their spin-isospin dependence as well as in the off-energy-shell properties, although all of them reproduce more or less the basic properties of two nucleon systems, both in the bound state (deuteron) and scattering states observed in experiments.
(One should note that experimental data of the two-body NN system give us only information of the on-energy shell because of the energy conservation of the two-body system.) 

\subsection{Saturation curves and stiffness of symmetric nuclear matter}

\begin{figure}[b]
\centering\includegraphics[width=2.5in,angle=0]{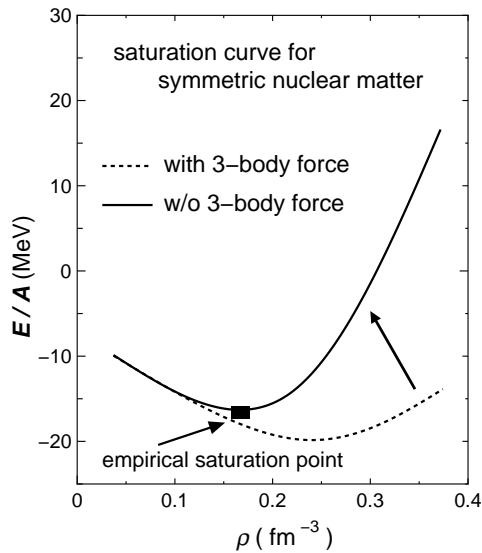}
\caption{An illustrative image of saturation curves in the symmetric nuclear matter (NM). The introduction of three-body forces makes the NM stiffer at high densities.}
\label{fig_SC}
\end{figure}
The single-particle energy in nuclear matter is calculated in the Brueckner-Hartree-Fock (BHF) framework with the use of the $G$-matrix interaction.
This is nothing but the binding energy of a nucleon in nuclear matter, that corresponds to the volume term of the binding energy per nucleon $\bar E = E/A$ in a finite nucleus.
The single-particle energy depends on the density $\rho$ of the nuclear matter and the curve of $\bar E(\rho)$ as a function of $\rho$ is called the {\em saturation curve} of the symmetric nuclear matter. 
(An illustrative example is shown in Fig.~\ref{fig_SC}. )
The saturation curve should have a minimum value $\bar E_0$ at a certain value of the density $\rho_0$ and the point $(\bar E_0, \rho_0)$ in the two-dimensional $\bar E - \rho$ plane is called the {\em saturation point} and  the values of $\bar E_0$ and $\rho_0$ are called as the {\em saturation energy} and the {\em saturation density} of the symmetric nuclear matter. 
The systematic observation of the r.m.s. radii of stable nuclei existing in nature suggests the saturation density to be about $\rho_0 \simeq 0.17$ fm$^{-3}$ and the volume term of the mass formula for finite nuclei gives the value of saturation energy of $\bar E_0 \simeq -16$ MeV.
Thus, the saturation point that should be reproduced by any satisfactory theories is the so-called {\em empirical saturation point}, ($\bar E_0$, $\rho_0$) = ($-16$ MeV, 0.17 fm$^{-3}$).

The history of the study of effective interactions and the saturation curve for symmetric nuclear matter tells us that it is impossible to reproduce the saturation point by any effective interactions evaluated from the non-relativistic $G$-matrix calculations starting from various kinds of NN-interaction models as long as only the two-body NN interaction is taken into account in the calculation.
Each model gives a saturation point apart from the empirical one and it is known that the evaluated saturation points lie along an almost straight line called the {\em Coester line}~\cite{COE70} that passes by the empirical saturation point with non-negligible amount of distance.  
Thus, it is one of the fundamental subjects in nuclear physics to reproduce the saturation point in symmetric nuclear matter.
A key is the role of the {\em many-body forces} in a dense nuclear matter.
As will be shown later, the introduction of three-body and four-body forces that act among three or four nucleons simultaneously at high-density nuclear matter can resolve the problem.
Within the non-relativistic framework, it is inevitable to introduce such many-body forces to reproduce the correct saturation point. 

In addition to the saturation point, the shape of the saturation curve is also important for understanding the dynamical properties of nuclear matter with the variation of its density. It is also an important subject to investigate the static and dynamic response of nuclear matter at high densities well above the saturation density ($\rho \gg \rho_0$) under the high pressure of gravity in high-density neutron stars.
The second derivative of the saturation curve gives a measure of the stiffness of the nuclear matter.
The incompressibility $K_{\rm SNM}$ of the nuclear matter is defined by the second derivative of $\bar{E}$ with respect to $\rho$ at the saturation density $\rho=\rho_0$, as $\displaystyle{ K_{\rm SNM} = 9 \rho_0^2 \frac{d^2 \bar{E}}{d \rho^2}\biggr|_{\rho_0} }$. 
A number of efforts have been made to fix the $K_{\rm SNM}$ value through various experimental information of finite nuclei or nuclear reactions, such as the measurements of excitation energies of giant monopole resonance (GMR) or high-energy heavy-ion (HI) collisions. 
Those experiments suggest rather large values of $K_{\rm SNM}$ range from about 200 MeV to 300 MeV.
Those values are much larger than those predicted by the $G$-matrix calculations with only the NN two-body force, since most of such $G$-matrix calculations give $K_{\rm SNM} \le$ 100 MeV.
This fact also suggests a need of introducing some additional repulsive effects, such as the three-body force of a repulsive nature, although the evaluated values of $K_{\rm SNM} \approx$ 200 MeV $\sim$ 300 MeV are far from conclusive for determining the precise stiffness of the symmetric nuclear matter and discussing the detail properties of such an additional repulsive effect.

Besides the $K_{\rm SNM}$ value, what is more important is the shape of the saturation curve itself, particularly at higher densities beyond the saturation density $\rho \gg \rho_0$, since the $K_{\rm SNM}$ gives only the curvature of the saturation curve at or near the saturation density $\rho \cong \rho_0$ by definition.
It only provides us an information about the stiffness of the symmetric nuclear matter near the saturation density.
Thus, it is quite important to obtain information about the stiffness of nuclear matter at densities much larger than the saturation value,  say, for example, $\rho > 2\rho_0 \sim 5\rho_0$. 
This is just the region where one expect to the high-density nuclear matter to be realized inside the neutron star, though it is the neutron matter rather than the symmetric nuclear matter.
It is one of the important goals of nuclear physics to predict reliable saturation curves both of symmetric nuclear matter as well as the neutron matter over the wide range of nucleon density including the high-density region of neutron-star matter.

In the following, we will see a recent challenge to determine the strength of the three-body and four-body force effects of repulsive nature through the precise analyses of nucleus-nucleus elastic scattering in laboratories.
The nucleus-nucleus interactions are derived microscopically from modern $G$-matrix effective NN interactions with the three-body and four-body force effects included. 
This makes it possible to directly discuss the stiffness of high-density nuclear matter as well as that of the neutron stars in the context of the effects of three-body and four-body forces that can be probed and tested by nucleus-nucleus elastic scattering  in laboratories on the Earth.

\section{Microscopic models for interaction between finite nuclei}

It is one of the important subjects in nuclear physics to understand the interactions between complex nuclei consisting of nucleons microscopically starting from the bare NN interaction in free space.
In constructing the nucleon-nucleus and nucleus-nucleus interactions starting from the bare NN interaction having a strong short-range repulsive core, one have to correctly handle the short-range correlation (SRC) originated from the short-range repulsive core of the NN interaction.
This is equivalent, in princile, to exactly solve the many-body dynamics in full model space up to extremely high excitation energies, that is almost impossible to perform in practice except for few-body systems consisting of nucleons at most $A\simeq 8$. 

One of the alternative ways of handling the SRC between interacting nucleons in nuclear many-body systems is to introduce an effective NN interaction that contains the SRC effectively in the multiple-scattering framework, such as the $T$-matrix, $R$-matrix and $G$-matrix interactions. 
The short-range singularities in the free-space NN interaction are smoothed out in those effective interactions.

\subsection{$G$-matrix folding model with local-density approximation}
The Brueckner's $G$-matrix is one of the simplest but reliable effective NN interaction that includes all the ladder diagrams of the NN interaction with the medium effect by surrounding nucleons in nuclear matter being taken into account.
This implies that the SRC induces the virtual excitations of the interacting NN pair into high-energy states outside the Fermi sphere and its effect is included, within the nuclear-matter approximation, in the resultant $G$-matrix effective NN interaction.
The $G$-matrix is then interpreted as an effective NN interaction in the ``cold'' or ``frozen'' nuclear matter of a uniform nucleon density.
This is the basis of the idea of the {\em folding models} discussed below for constructing nucleon-nucleus or nucleus-nucleus interaction potentials in terms of the $G$-matrix.
In the folding model, the $G$-matrix is folded over the nucleon-density distributions of the interacting finite nuclear systems in their {\em ground states}.

Since the $G$-matrix $v_{\rm NN}$ is evaluated in the nuclear matter of a uniform constant density, the evaluated $G$-matrix depends on the nucleon density $\rho$ of the nuclear matter.
Because of the finite size of the nucleus, the environment, or the {\em local density}, for an interacting nucleon pair embedded in finite nuclear systems changes rapidly with the spatial positions where the nucleon pair exist. 
Thus, the folding model adopts the {\em local density approximation} (LDA), in which the uniform density of the nuclear matter for evaluating the $G$-matrix is assumed to be the same as the local density at the spatial positions of the interacting nucleons in finite nuclei.
In other words, we use the different strength of the $G$-matrix depending on the local density at the positions of interacting nucleons in finite nuclei in calculating the nuclear potential in the folding model.
The G-matrix also depends on the starting energy $\omega$, that corresponds to the incident energy per nucleon, $E/A$, in the case of the nucleus-nucleus scattering.

\subsection{Optical potential and folding model with complex $G$-matrix} 
The $G$-matrix is real for bound states because both interacting nucleons are bound inside the Fermi sphere in the initial and finial states of interaction, although they can be excited virtually to continuum states outside the Fermi sphere in the intermediate stages of the interaction.
This kind of real $G$-matrix are widely used in various nuclear-structure calculations.
On the other hand, the $G$-matrix derived under the scattering boundary condition becomes complex because both of the interacting nucleons can be excited to continuum (scattering) states outside the Fermi sphere in the final states of the interaction.
In the nucleus-nucleus scattering, for example, the interacting nucleon pair, one in the projectile nucleus and the other in the target one, is in their relative scattering states. 
Thus, the $G$-matrix becomes complex and the complex $G$-matrix should be used in constructing the nucleus-nucleus potential microscopically by the folding model.
In fact, the nucleon-nucleus or nucleus-nucleus potential evaluated phenomenologically from the elastic-scattering experiment is known to be a complex potential called the {\em optical potential}.
The optical potential is a complex potential having real and imaginary parts, each of which is assumed to have a simple geometrical functional form, such as the three-parameter {\em Woods-Saxson form}; $f(R)= -V_0/[1+\exp{((R-R_0)/a_0)}]$, 
and the parameter values of the real and imaginary parts are to be optimized so that the angular distribution of the elastic-scattering cross section calculated with the optical potential reproduces that of the experimentally observed elastic-scattering cross section.

The imaginary part of the optical potential represents the loss of flux from the elastic-scattering channel.
In the nucleon-nucleus or nucleus-nucleus collision, many kinds of nuclear reactions take place and some fraction of the incident flux, that initially comes into the entrance channel (i.e. to the elastic-scattering channel), escapes from the entrance elastic-scattering channel to non-elastic reaction channels, such as inelastic scattering and various kinds of rearrangement, breakup and fusion reactions, etc. 
The flux loss can be represented approximately by the imaginary part of the interaction potential.

The complex folding-model potential obtained from the complex $G$-matrix is a promising candidate of the microscopic interaction model for explaining the {\em complex optical potentials} derived phenomenologically from the nucleon-nucleus or nucleus-nucleus elastic-scattering experiments, that will be shown below in some detail.

In the following sections, we will concentrate our discussion to the {\em double-folding model} (DFM) for constructing interaction potentials for nucleus-nucleus systems.
This is because the nucleus-nucleus scattering is one of the ideal tools for probing the saturation property of nuclear matter at high densities beyond the saturation density, say up to twice the normal density $\rho\sim 2\rho_0$.
Combining the theoretical analyses of the nucleus-nucleus scattering by the DFM and the BHF study of the saturation curves of the nuclear matter and the neutron-star matter using a common $G$-matrix, it will be made possible to determine unknown strength of the three-body force (or many-body forces, in general) very accurately and to discuss directly the so-called ``maximum-mass problem on neutron stars'' together with the stiffness of nuclear matter and neutron stars.

\section{Double-folding model with $G$-matrix interactions}

In the case of the nucleus-nucleus system, the folding-model potential can be written as a Hartree-Fock type potential; 
\begin{equation}
U_{\rm{F}} = \sum_{i\in P, \,  j\in T}{[<ij|v_{\rm{D}}|ij>+<ij|v_{\rm{EX}}|ji>]} \;
\equiv\; U_{\rm{D}}+U_{\rm{EX}}, \label{eq:hfp}
\end{equation}
where $v_{\rm{D}}$ and $v_{\rm{EX}}$ are the direct and exchange parts of the $G$-matrix $v_{\rm{NN}}$, while $<ij|v_{\rm{D}}|ij>$ and $<ij|v_{\rm{EX}}|ij>$ represent the matrix elements of the direct and knock-on exchange parts, respectively, of the $G$-matrix interaction.
Here, $|ij>$ denotes the direct product of the ground-state wave functions 
$\Psi_{\rm P}$ and $ \Psi_{\rm T}$ of the interacting nuclei; 
i.e.~$|ij> = | \Psi_{\rm P}\otimes \Psi_{\rm T}>$.

The microscopic nucleus-nucleus potential of the Hartree-Fock type potential with the $G$-matrix effective NN interaction defined in Eq.~\ref{eq:hfp}  is equivalent to the double-folding-model (DFM) potential with the density-dependent effective  interaction.

\begin{figure}[t]
\centering\includegraphics[width=2.5in,angle=0]{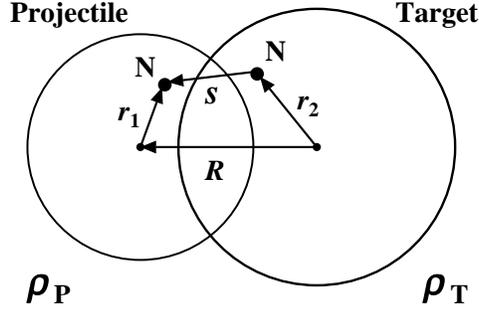}
\caption{Definitions of the coordinates appearing the Eq.~\ref{eq:dfmd} for calculating the direct part of the DFM potential between projectile nucleus (P) and the target one (T).}
\label{fig_DFcood}
\end{figure}

The direct part of the Hartree-Fock type potential is rewritten in the double-folding type integration where the direct part of the $G$-matrix $v_{\rm D}$ is doubly folded over the nucleon densities of the interacting nuclei, P (projectile) and T (target), as
\begin{equation}
U_{\rm{D}}(\bm{R})=\int{\rho_{\rm\: P}(\bm{r}_{1}) \rho_{\rm\: T}(\bm{r}_{2}) 
v_{\rm\, D}(\bm{s}; \rho, E/A)d\bm{r}_{1} d\bm{r}_{2}}. \label{eq:dfmd}
\end{equation}
Here, $\bm{R}$ is the relative coordinate between the centers of mass of the interacting nuclei P and T and $\bm{s}\equiv \bm{R}+\bm{r}_1-\bm{r}_2$ is the relative coordinate between a nucleon at the spatial position $\bm{r}_1$ with respect to the center of mass (c.m.) of the nucleus P and another nucleon at the spatial position $\bm{r}_2$ with respect to the c.m. of the nucleus T.
They interact with each other though the direct part of the $G$-matrix, $v_{\rm\, D}$ (see Fig.~\ref{fig_DFcood}).
The nucleon densities of the nuclei P and T are defined as
\begin{eqnarray}
\rho_{\rm\: P}(\bm{r}_{1}) &=& \Bigl< \Psi_{\rm P} \Bigm| 
\sum_{i \in {\rm P}}^{A_{\rm P}} \delta ( \bm{r}_1 - \bm{r}_i ) 
\Bigm| \Psi_{\rm P} \Bigr> \; , \\
\rho_{\rm\: T}(\bm{r}_{2}) &=& \Bigl< \Psi_{\rm T} \Bigm| 
\sum_{j \in {\rm T}}^{A_{\rm T}} \delta ( \bm{r}_2 - \bm{r}_j ) 
\Bigm| \Psi_{\rm T} \Bigr>\; ,
\end{eqnarray}
where $A_{\rm P}$ and $A_{\rm T}$ are the mass numbers of the nucleus P and the nucleus T, respectively.
In Eq.~\ref{eq:dfmd}, the direct part of the $G$-matrix $v_{\rm\, D}$ depends on the local density $\rho$.
In the present DFM calculation, the local density is defined by the {\em sum of the densities} of P and T, as $\rho = \rho_{\rm P} + \rho_{\rm T}$ evaluated at the individual positions or the mid-point of the interacting nucleons, as will be discussed in more detail in the following subsection. 
This prescription for evaluating the local density is called as the {\em frozen-density approximation} (FDA) and widely used in most DFM calculations~\cite{FAR85,KAT02,FUR06,FUR09}.
The FDA is also used in evaluating the local density in calculating the exchange part below.

The exchange part of the DFM potential is a non-local potential in principle but is known to be approximated in good accuracy by the following local-potential form of the so-called ``Brieva-Rook type''~\cite{BR77a,BR77b,BR78};  
\begin{eqnarray}
U_{\rm{EX}}(\bm{R})&=&\int{\rho_{P}(\bm{r}_1, \bm{r}_1+\bm{s}) \rho_{T}(\bm{r}_2, \bm{r}_2-\bm{s}) v_{\rm{EX}}(s; \rho, E/A)} \nonumber \\
&&\times \exp{ \left[ \frac{i\bm{k}(R)\cdot \bm{s}}{M} \right] } d\bm{r}_1 d\bm{r}_2. 
\label{eq:dfexchange}
\end{eqnarray}
Here, $\bm{k}(R)$ is the local momentum for the nucleus-nucleus relative motion defined by 
\begin{equation}
k^2(R)=\frac{2mM}{\hbar^2}[E_{\rm{c.m.}}-{\rm{Re}}U_{\rm{F}}(R)-V_{\rm{coul}}(R)], \label{eq:kkk}
\end{equation}
where, $M=A_{\rm P} A_{\rm T}/(A_{\rm P} + A_{\rm T})$, $E_{\rm{c.m.}}$ is the center-of-mass energy, $E/A$ is the incident energy per nucleon, $m$ is the nucleon mass and $V_{\rm{coul}}$ is the Coulomb potential. 

The exchange part is calculated self-consistently on the basis of the local-energy approximation through Eq.~(\ref{eq:kkk}). 
Here, the Coulomb potential $V_{\rm{coul}}$ is also obtained by folding the NN Coulomb potential with the proton density distributions of the projectile and target nuclei. 
The density matrix $\rho(\bm{r}, \bm{r}')$ is approximated in the same manner as in \cite{NEG72}; 
\begin{equation}
\rho (\bm{r}, \bm{r}')
=\frac{3}{k^{\rm{eff}}_{\rm{F}}\cdot s }\: j_{1}(k^{\rm{eff}}_{\rm{F}}\cdot s)\: 
\rho \Big(\frac{\bm{r}+\bm{r}'}{2}\Big), 
\label{eq:exchden}
\end{equation}
where $k^{\rm{eff}}_{\rm{F}}$ is the effective Fermi momentum \cite{CAM78} defined by
\begin{equation}
k^{\rm{eff}}_{\rm{F}} 
=\Big[ (3\pi^2 \rho )^{2/3}+\frac{5C_{\rm{s}}[\nabla\rho^2]}{3\rho^2}
+\frac{5\nabla ^2\rho}{36\rho} \Big]^{1/2}, \;\; 
\label{eq:kf}
\end{equation}
where $C_{\rm{s}} = 1/4$ following Ref.~\cite{KHO01}. 
The detail methods for calculating $U_{\rm{D}}$ (direct part)  and  $U_{\rm{EX}}$ (exchange part) are the same as those given 
in Ref.~\cite{NAG85} and \cite{KHO94}, respectively. 

\subsection{Frozen-density approximation}

One have to define the local density $\rho$ at which the strength of the $G$-matrix is evaluated in the DFM calculations.
In the nucleus-nucleus system, there are several prescriptions of defining the local density $\rho$ that defines the nuclear environment for the interacting nucleon pair embedded in the colliding nucleus-nucleus system. 

One is the {\em average} prescription, in which the local density is defined by the average,  either the geometric average or the arithmetic one, of the local densities corresponding to the individual positions of the interacting nucleon pair;
\begin{equation}
\rho = \frac12 \bigm( \rho_{\rm P}(\bm r_1) + \rho_{\rm T}(\bm r_2) \bigm) \;\;\;\;\;\;
\mbox{or} \;\;\;\;\;\;
\rho = \sqrt{ \rho_{\rm P}(\bm r_1)\: \rho_{\rm T}(\bm r_2)}\; ,
\label{eq:Ave}
\end{equation}
or, alternatively, the average of the local densities corresponding to the mid-point of the positions of the interacting nucleon pair.

Another prescription, that is widely adopted in most DFM calculations, is the {\em sum} of the local densities corresponding to the individual positions of the interacting nucleon pair 
\begin{eqnarray}
\rho &=& \rho_{\rm P}(\bm r_1) + \rho_{\rm T}(\bm r_2)\; , \label{eq:FDA1} 
 \label{eq:FDA2}
\end{eqnarray}
or, alternatively, to the mid-point of the nucleons;
\begin{eqnarray}
\rho &=& \rho_{\rm P}\bigl(\bm{r}_1- \textstyle{\frac{1}{2}}\bm{s} \bigr) 
       + \rho_{\rm T}\bigl(\bm{r}_2 + \textstyle{\frac{1}{2}}\bm{s} \bigr)\; , \label{eq:FDA3}
\end{eqnarray}
both of which are called the {\em frozen-density approximation} (FDA).
In FDA, the interacting nucleon pair is assumed to feel the local density defined as the sum of densities of colliding nuclei.

The difference between the two prescriptions, the ``average'' and the ``sum'' (FDA), become significant when two nuclei come close to each other with a substantial overlap of the two density distributions. 
In the case of FDA, the local density can reach to the sum of the central densities of the interacting nuclei that is close to twice the saturation density, 
say $\rho \approx \rho_{\rm P}(0)+\rho_{\rm T}(0) \approx 2\rho_0$.
On the other hand, the local density does not exceed the saturation density  $\rho \leq \rho_0$ in the average prescription.

One might postulate the {\em average prescription} to be more reasonable than the FDA because of the saturation of nucleon density in finite nuclei and might require that the local density should  not exceed the saturation density $\rho \leq \rho_0$ even when two nuclei come close to each other at very short distances.
However, the strict condition that the local density $\rho$ should not exceed the saturation density $\rho_0$ of the ``stable'' nuclei implies that two stable nuclei in their ground states cannot heavily overlap to each other in coordinate space. 
Otherwise either or both of the interacting nuclei are forced to be excited to highly excited states to fulfill the Pauli Principle.
In other words, the strict condition, $\rho \leq \rho_0$, inevitably leads to a strong rearrangements of nucleon configurations among the two interacting nuclei when they come close to short distances.
This implies that most of the incident flux is removed from the elastic-scattering channel to non-elastic reaction channels. 
In practice, these kinds of dynamic reaction processes may occur with high probabilities in high-energy HI collisions and almost all incident flux will be lost from the elastic-scattering channel.

Such kind of hard collision, where the incident flux is completely removed from the elastic channel to the reaction channels, corresponds to the so-called {\em shadow scattering} that is the quantum-mechanical elastic scattering by a black-body target. 
In the case of shadow-scattering, all the information about the internal region of the interaction potential between colliding nuclei are completely lost and no refractive scattering due to the attractive real potential should not be observed in elastic scattering.
In fact, in many cases, the elastic scattering between complex nuclei are dominated by the strong absorption due to a huge number of non-elastic reactions occurring, when two nuclei come close to each other, and the elastic-scattering cross section often shows a typical characteristics of the strong absorption.
In such cases, there is no hope to get any information about interaction potentials at short distances and the prescriptions for defining the local density, either average or sum, are not relevant to the elastic-scattering observables. 

However, it is often the case that experimental data for elastic scattering of light HIs, such as the $^{16}$O + $^{16}$O and $^{12}$C+$^{12}$C systems at intermediate energies, show characteristic angular distributions typical to the refractive scattering generated by the attractive real potential at short distances that strongly participates in the elastic scattering cross section at large scattering angles~\cite{BOH93}.
This indicates that the HI elastic scattering can probe an important information about the strength of interaction potential at very short distances~\cite{KONDO90,BOH93} where two nuclei, both kept staying in their ground states, come close to each other at very short distances.
This implies that there exist non-negligible probability of two interacting nuclei being kept in the ground state at short relative distances~\cite{DNR}, where the local density should be a simple sum of the undisturbed densities of the individual nuclei in their ground states that can reach to twice the saturation density $\rho \sim 2\rho_0$.

Of course, it is natural to expect that major part of the incident flux are removed from the elastic-scattering channel to various non-elastic reaction channels.
However, the refractive scattering widely observed in experiments clearly tells us that the incident flux is not completely removed from the elastic-scattering channel and there remains, though only a tiny fraction, the elastic-scattering component of the wave function for relative motion between the interacting nuclei at very short distances. 
The elastic-scattering experiments can probe such a tiny but non-negligible component of high-local density region that carries a quite valuable information about the high-density nuclear matter.

Nowadays, the superiority of FDA (defined by Eqs.~(\ref{eq:FDA2}) and (\ref{eq:FDA3})) over the average prescription (defined by Eqs.~(\ref{eq:Ave})) has been made clear through experimental and theoretical analyses of a number of such refractive scattering of light HI systems~\cite{KHO94, KHO97, KHO01, KAT02, SAT79}.
The average prescription largely overestimates the strength of the real potential at short distance and one needs to introduce an artificial renormalization factor for strongly reducing the real-potential strength so evaluated in order to reproduce the elastic-scattering experiments.
On the other hand, no renormalization to the real potential evaluated with FDA is essentially required.
Thus, it is widely recognized that FDA is the best prescription for defining the local density to be used in the DFM calculation of HI potentials, as long as one uses the reliable, realistic density-dependent effective interactions~\cite{DUG81, CAR96, KHO01, FUR06,FUR09} such as the $G$-matrices~\cite{FUR09}.
Therefore,  we adopt the FDA in the calculation of DFM potential with the $G$-matrix throughout the following discussions.

\section{A brief history of DFM studies of the HI scattering}

The first successful DFM for constructing the nucleus-nucleus potentials was the M3Y folding model proposed by Satchler and Love in 1970s and its achievements in those days were summarized in a review article~\cite{SAT79}. 
The M3Y (Michigan 3-range Yukawa) interaction~\cite{M3Y77} is a kind of $G$-matrix represented by a linear combination of the three-range Yukawa functions.
It has no density dependence and has only the real part.
Thus, the M3Y-folding model only generates the real part of the nucleus-nucleus potentials.
Despite the simpleness of the M3Y interaction, the DFM with the M3Y interaction was quite successful in the application to analyzing the low-energy HI scattering and marked a milestone in the study of microscopic theory for nucleus-nucleus interaction in those days.

However, it was soon recognized that the low-energy HI scattering was only sensitive to the surface part of the potential in most cases. 
The ``success'' of the M3Y folding model thus meant that it only gave a reasonable strength of the real part of HI potentials around the spatial region where two nuclei just touch their surfaces keeping a large relative distance between their centers of mass.
This implies that no meaningful information about the strength and shape of the potential at short distances was obtained by the limited success of the M3Y folding model.

Since then, it took 30 years to mark a really important another milestone in the study of microscopic theory for predicting reliable {\em complex} nucleus-nucleus interaction, that was marked by Furumoto, Yamamoto and the present author in 2008 $\sim$ 2009 \cite{FUR08,FUR09R,FUR09}.
The new interaction model has been quite successful in reproducing almost all existing data for HI scattering at incident energies per nucleon of $E/A \approx 30 \sim 200$ MeV~\cite{FUR09,FUR12GP} and can be applied to any HI scattering system over the wide range of incident energies up to $E/A \approx 400$ MeV and any combinations of projectile and target nuclei in the nuclear chart as long as the nucleon density is available.
For the sake of further global use of the successful interaction model, they have proposed a {\em global optical potential} based on the microscopic folding model that is applicable to any combinations of the nucleus-nucleus systems at any incident energy within the range of  $E/A = 30 \sim 400$ MeV~\cite{FUR12GP}. 

What is more important is that they unveil the decisive roles of the many-body forces, particularly of the three-body force of repulsive nature.
The many-body forces play important roles both in predicting the strength and the shape of the nucleus-nucleus potential at short relative distances as well as in determining the stiffness of nuclear matter, namely the shape of the saturation curve particularly in the high-density side beyond the saturation density 
(cf. Fig.\ref{fig_SC}).

Before presenting the latest achievements, we briefly review a history of the development of the folding-model studies of HI scattering in some detail.

\subsection{M3Y and DDM3Y interactions}
%
\subsubsection{M3Y interaction}
The M3Y interaction is a real interaction having no imaginary part and has a very weak energy dependence.
Hence, the DFM potential calculated with the M3Y interaction is a real potential and it has to be supplemented by a phenomenological imaginary potential when applied to the analyses of HI scattering experiments~\cite{SAT79}. 

The M3Y interaction is composed of a sum of three-range Yukawa potentials.
The strengths of the individual range of the Yukawa potential was fixed so that the matrix elements of the M3Y interaction reproduced the real-part strength of the original $G$-matrix evaluated at a certain low value of the nuclear matter density, say about one third of the saturation density $\rho \approx \frac13\rho_0$. 
The M3Y is the real potential  because it was originally designed for the application to nuclear-structure studies and later it was applied to the construction of the real part of nucleus-nucleus potentials to be used in the study of HI scattering.
More over, the exchange part of the simplest version of the M3Y interaction is approximated by a zero-range potential of the $\delta$-function form that makes the DFM calculation of the exchange part extremely simple as in the case of the direct part.

These simplifications make the originally complicated numerical calculation of the DFM potentials (Eqs.~(\ref{eq:dfmd}) $\sim$ (\ref{eq:kf})) quite light and easily accessible by anyone who was not an expert in this research area.
Despite a drastic simplification, the DFM with the M3Y interaction has been quite successful in applications to the HI scattering at relatively lower energies, say the incident energy per nucleon $E/A$ being less than about 25 $\sim$ 30 MeV~\cite{SAT79}.

\begin{figure}[b]
\centering\includegraphics[width=2.8in,angle=0]{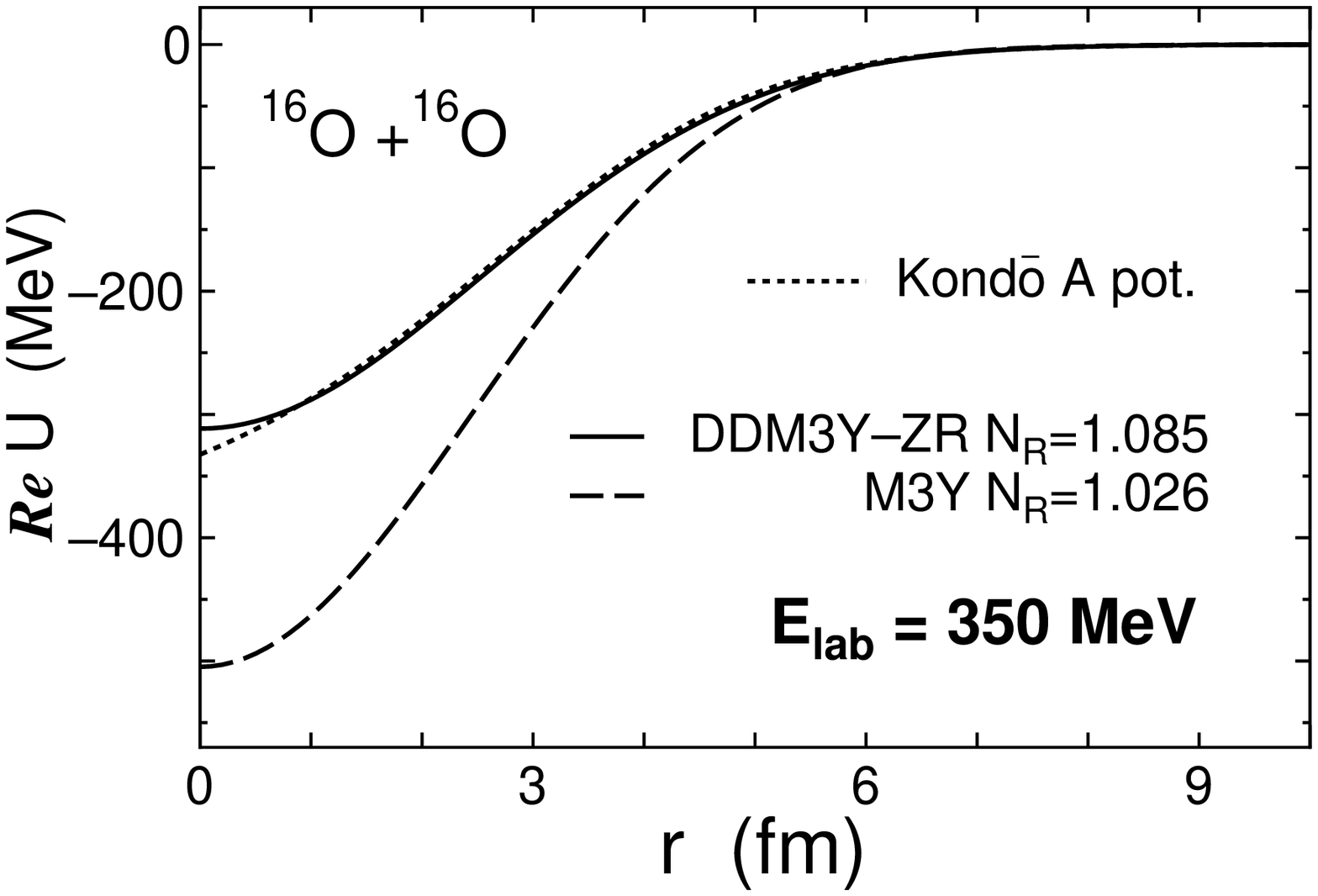}
\hspace{10mm}
\centering\includegraphics[width=2.5in,angle=0]{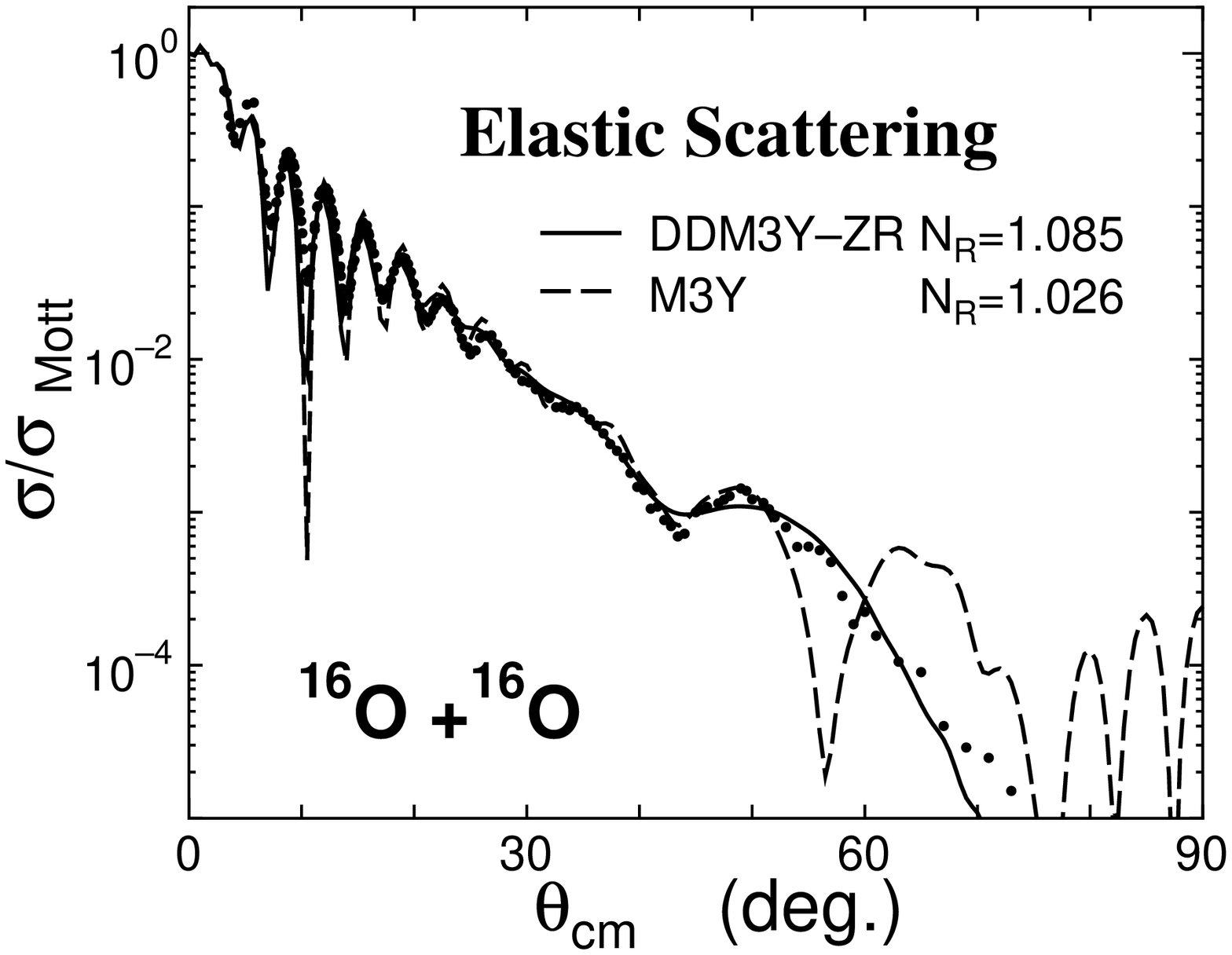}
\caption{(Left): The DFM real potential for the $^{16}$O + $^{16}$O scattering at $E_{\rm lab}$ = 350 MeV calculated with the use of the M3Y (dashed curve) and DDM3Y (solid curve) interactions, that is compared with the phenomenological potential (Kondo-A potential: dotted curve). (Right): The cross section for $^{16}$O + $^{16}$O elastic scattering at $E_{\rm lab}$ = 350 MeV calculated by the DFM potential with the M3Y (dashed curve) and DDM3Y (solid curve) interactions, compared with the experimental data. 
The figures are taken from Ref.~\cite{KAT02}    }
\label{fig_DFM}
\end{figure}

Figure~\ref{fig_DFM} show an example of the DFM potential calculated with the M3Y interaction and its application to the analysis of elastic scattering of the  $^{16}$O + $^{16}$O system at $E_{\rm lab}=350$ MeV ($E/A \simeq 22$ MeV).
The dashed curve on the left panel shows the real DFM potential calculated with the M3Y interaction.
The potential is multiplied by a renormalization factor of 1.026, just a 2.6\% modification of the original strength, that is chosen to make a fine tuning of fit to the data. 
The result is shown by the dashed curve in the right panel~\cite{KAT02}.

It is seen that the M3Y interaction produces very deep real potential, its central depth reaching to 500 MeV. 
The real potential obtained by the M3Y folding supplemented by a phenomenological imaginary potential (though not displayed in the figure) produces the elastic-scattering cross section that reproduces the experimental data quite well except for those at extremely large scattering angles $\theta_{\rm c.m.}\geq 50^\circ$.
The characteristic shape of the observed angular distribution of the experimental data, a rapid oscillation at small angles followed by a smooth fall off up to $\sim 50^\circ$ is well reproduced by the calculation (the dashed curve).
The same quality of fits to other experimental data have also been achieved in low-energy HI scattering based of the DFM potential derived from the M3Y interaction~\cite{SAT79}. 
The modifications of the DFM potential strength required to attain the best fits to the data are known to be of the order of $\pm 10 \%$, that may indicate an acceptable level of ``success'' of the model~\cite{SAT79}. 

\subsubsection{DDM3Y interaction}
It is known that the low-energy HI scattering, in most cases, is dominated by the strong absorption of incident flux to a huge number of non-elastic reaction channels, as mentioned before, and the elastic-scattering cross sections at most scattering angles are not sensitive to the depth or shape of the real potential at short distances less than the grazing distance (or so-called strong-absorption radius).
Therefore, it is rather difficult to probe the short-range part of the HI potential by the low-energy HI scattering.

For example, the calculated cross sections shown by the solid curve in the right panel of Fig.~\ref{fig_DFM} are obtained with the potential shown by the solid curve in the left panel (labeled by ``DDM3Y-ZR'') that is much shallower than the potential given by the M3Y interaction (shown by the dashed curve), particularly at short distances.

The shallower potential is obtained by the so-called {\em DDM3Y} interaction~\cite{KOB82,FAR85} that is a density-dependent (DD) version of the M3Y interaction obtained from the original M3Y interaction $g(s, \bar E)$ multiplied by a density-dependent factor $f(\rho, \bar E)$ as 
\begin{equation}
v_{\rm DDM3Y}(s, \rho, \bar E)=g(s, \bar E)\: f(\rho, \bar E).
\label{eq:DDM3Y1}
\end{equation}
Here $\bar E \equiv E/A$.
The DDM3Y interaction itself is not directly derived from any realistic $G$-matrix but the density-dependent factor is taken to be a suitable functional form, such as 
\begin{equation}
f(\rho, \bar E) = C(\bar E)\: \bigm( 1 + \alpha (\bar E)\: {\rm e}^{-\beta(\bar E) \rho} \bigm).
\label{eq:DDM3Y2}
\end{equation}
The parameters $C(\bar E)$, $\alpha (\bar E)$, $\beta(\bar E)$ of the density-dependent factor were determined at each energy so that the volume integral of $v_{\rm DDM3Y}(s, \rho, \bar E)$ matches the volume integral of the real part of a suitable $G$-matrix for various density $\rho$ of the nuclear matter~\cite{FAR85}.
The local density $\rho$ is evaluated in the prescription of FDA as $\rho=\rho_{\rm P}+\rho_{\rm T}$.
The effect of density dependence is quite significant at small distances between the interacting nuclei, as seen by the large reduction from the dashed curve (M3Y) to the solid one (DDM3Y) in the left panel of Fig.~\ref{fig_DFM}.

It is now surprising that the two very different potentials give almost the same cross sections over the entire range of scattering angles where the experimental data exist.
The difference only comes to appear at the largest scattering angles over $\theta_{\rm c.m.} \approx50^\circ$ where the solid curve keeps following the experimental data points while the dashed curve does not.
It should be noted that the potential derived from the DDM3Y interaction becomes close to that derived from the M3Y around the nuclear surface region $R \geq 6$ fm.
This implies that the cross sections at small and middle angles up to  $\theta_{\rm c.m.} \approx50^\circ$ is only sensitive to the tail part ($R \geq 6$ fm in the present case) of the real potential.
In other words, the potential depth inside the nuclear surface, say $R \leq 5$ fm is not relevant at all to generating the characteristic angular distribution at angles less than  $\theta_{\rm c.m.} \approx50^\circ$.
Namely, only cross sections at extremely backward angles ($\theta_{\rm c.m.} \geq 50^\circ$ in the present case) can probe the spatial region where two potentials have significant difference shown in the left panel of Fig.~\ref{fig_DFM}.

The scattering of light heavy ions at intermediate energies above $E/A \geq 50\sim 100$ MeV becomes rather transparent and displays a refractive nature generated by the strongly attractive real potential at short distances. 
The HI scattering at those energies becomes very sensitive to the shape and strength of the real potential at small separation where two nuclei strongly overlap to each other.
In such situations, the density dependence of the effective interaction becomes important  and the scattering experiments may be able to probe meaningful information about the high-density nuclear matter.

\subsection{DFM with complex $G$-matrix: CEG07 and MP interactions}

The DDM3Y interaction was first applied to the nuclear rainbow phenomena in $\alpha$-particle scattering~\cite{KOB82}, being the typical refractive scattering by the effects of attractive real potentials, and later it has widely been applied to various kinds of the HI scattering~\cite{BRAN97} with considerable success.
However, the DDM3Y interaction has only the real part and the generated DFM potential becomes also real. 
Hence, the real DFM potential has to be supplemented by a phenomenological imaginary potential in practical analyses of HI-scattering experiments as in the case of M3Y.
Moreover, another serious problem is that the density dependence of the DDM3Y interaction is just introduced by hand as a density-dependent factor $f(\rho, \bar E)$ which is simply multiplied by the original M3Y interaction as in Eqs.~(\ref{eq:DDM3Y1}) and (\ref{eq:DDM3Y2}) and has no fundamental theoretical background, particularly in the high-density domain above the saturation density.
For a correct understanding of the nucleus-nucleus interactions to be used in nuclear-reaction studies from the microscopic view point, it is strongly expected to construct a reliable complex $G$-matrix having a density dependence that properly represents the medium effects in nuclear matter for nucleon densities up to twice the saturation density.

Some attempts to construct the complex $G$-matrix to be used in nuclear reaction studies were made in late 70s to early 80s by several groups.
The typical one is the work by Oxford group led by Brieva and Rook~\cite{BR77a,BR77b,BR78} and another is the work by Miyazaki group led by Nagata~\cite{CEG83,CEG86}.
Individual groups constructed their original complex and density-dependent $G$-matrix effective NN interactions to be used in calculating the complex optical potential for nucleon-nucleus elastic scattering by the folding model.
They were successfully applied to the proton-nucleus elastic scattering at low to intermediate energies.
Those effective interactions, however, were only applicable to nucleon-nucleus systems because they were evaluated at nuclear-matter density only up to the saturation value $\rho_0$ and, hence, could not be applicable to nucleus-nucleus systems.
Another problem was that those effective interactions were generated by a very prototype NN-interaction models in free space, such as the Hamada-Jonston type potential~\cite{HJ62}.
This resulted in too small binding energies in nuclear matter less than one half of the empirical value of $\bar E \cong -16$ MeV in the BHF calculation with those $G$-matrices.

To overcome those problems in the pioneering works, 
an ideal complex $G$-matrix has recently be proposed by Furumoto, Yamamoto and the present author.
They have marked another important milestone and have established one of the most reliable microscopic interaction models to date for nucleus-nucleus reaction systems on the firm theoretical base starting from the free-space NN interaction.
It is a complex, density-dependent $G$-matrix evaluated by a latest elaborated NN interaction model in free space.
The $G$-matrix is applicable to the local density up to twice the saturation density and to the HI reactions at the energy per nucleon $E/A$ ranging  from $30\sim 400$ MeV.

\begin{figure}[b]
\centering\includegraphics[width=2.5in]{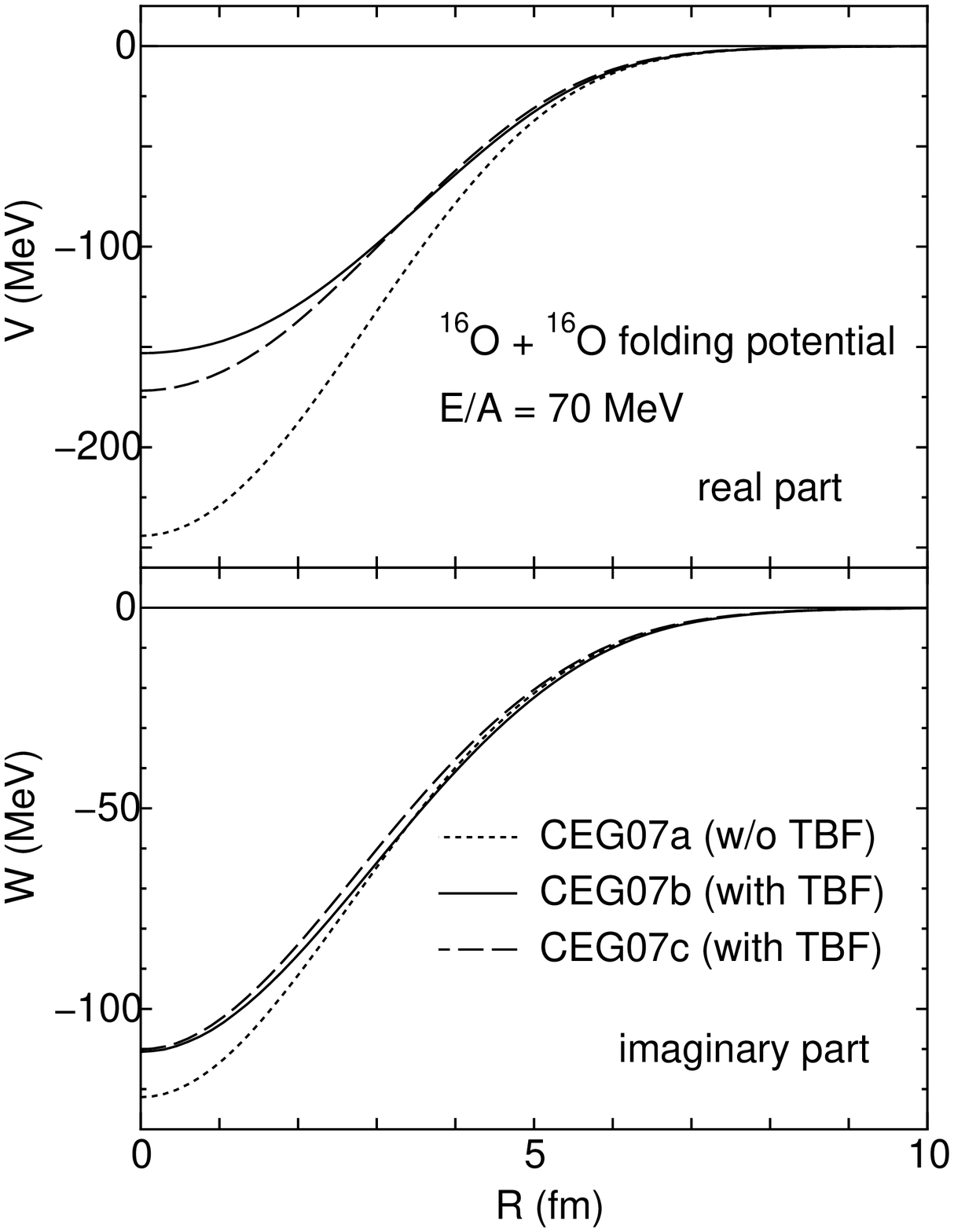}\hspace{15mm}\centering\includegraphics[width=2.5in]{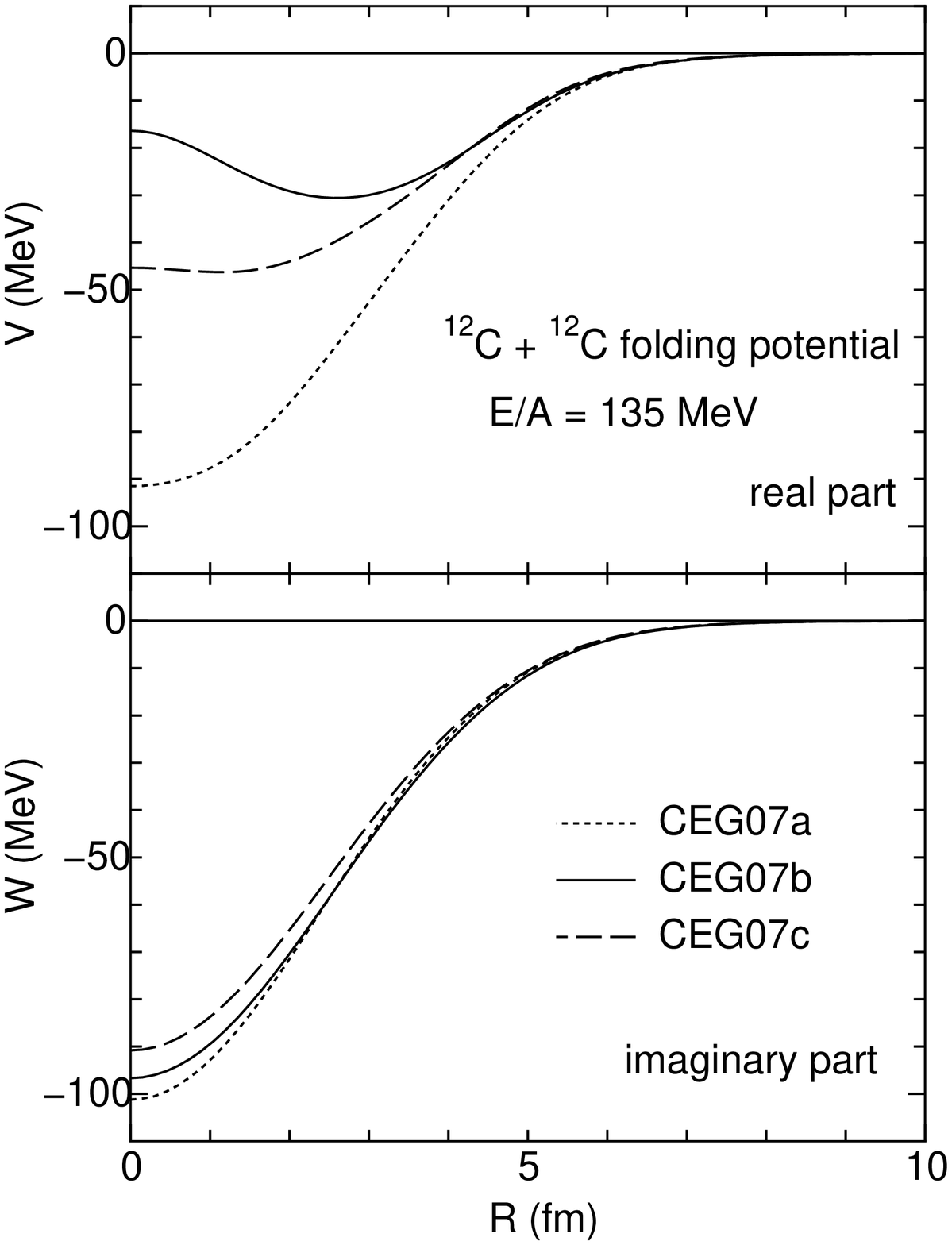}
\caption{
The double-folding-model (DFP) potentials for elastic scattering of the $^{16}$O +  $^{16}$O system at $E/A$ = 70 MeV (left) and of the $^{12}$C + $^{12}$C one at $E/A$ = 135 MeV (right), respectively. 
The dotted curves show the result without the 3NF effect (CEG07a), while the solid (CEG07b) and dashed (CEG07c) curves are the results with the 3NF effect. 
The figures are taken from Ref.~\cite{FUR09} .
}
\label{fig:CEG07POT}
\end{figure}

\begin{figure}[t]
\centering\includegraphics[width=2.5in]{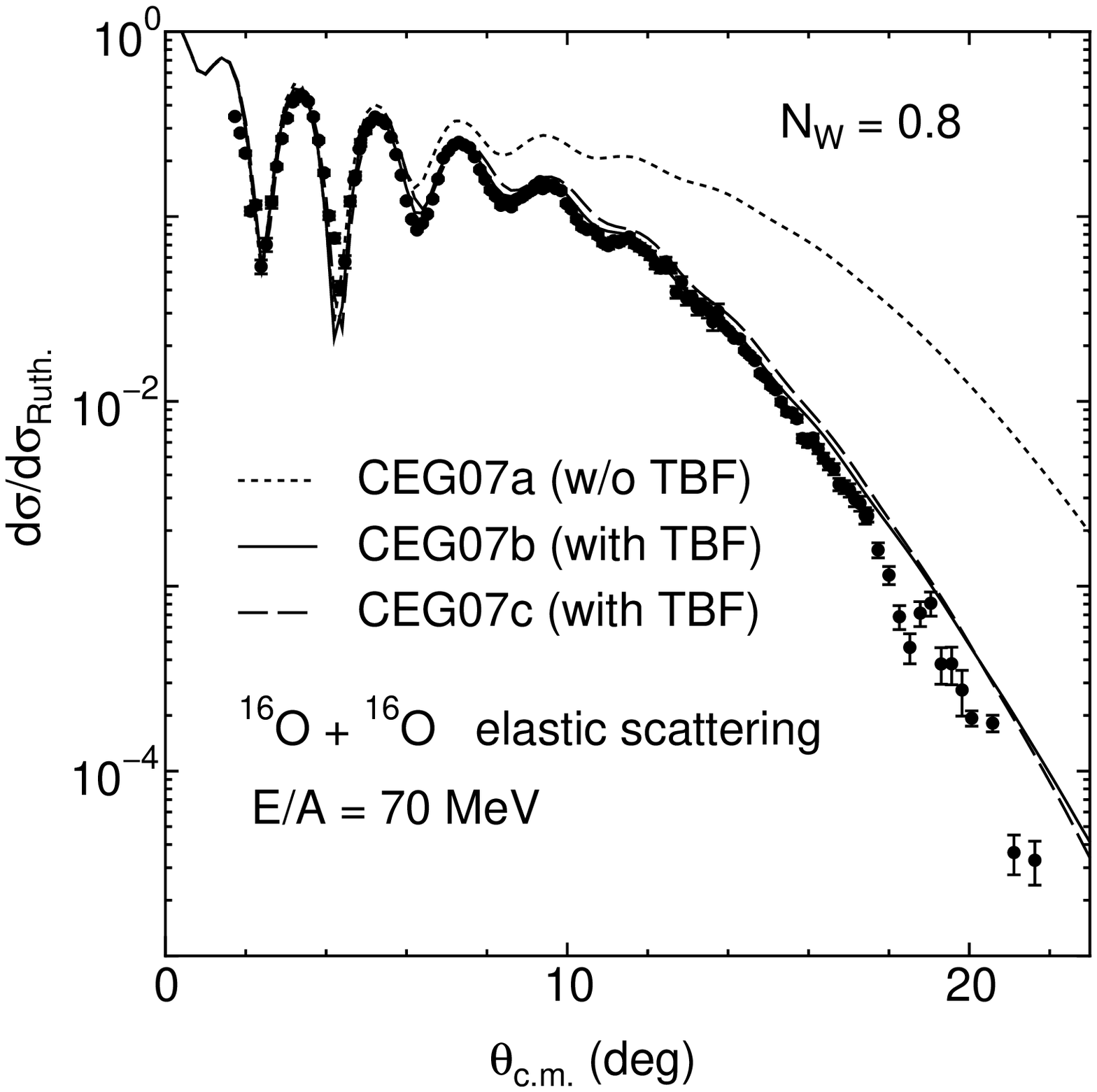}\hspace{15mm}\centering\includegraphics[width=2.5in]{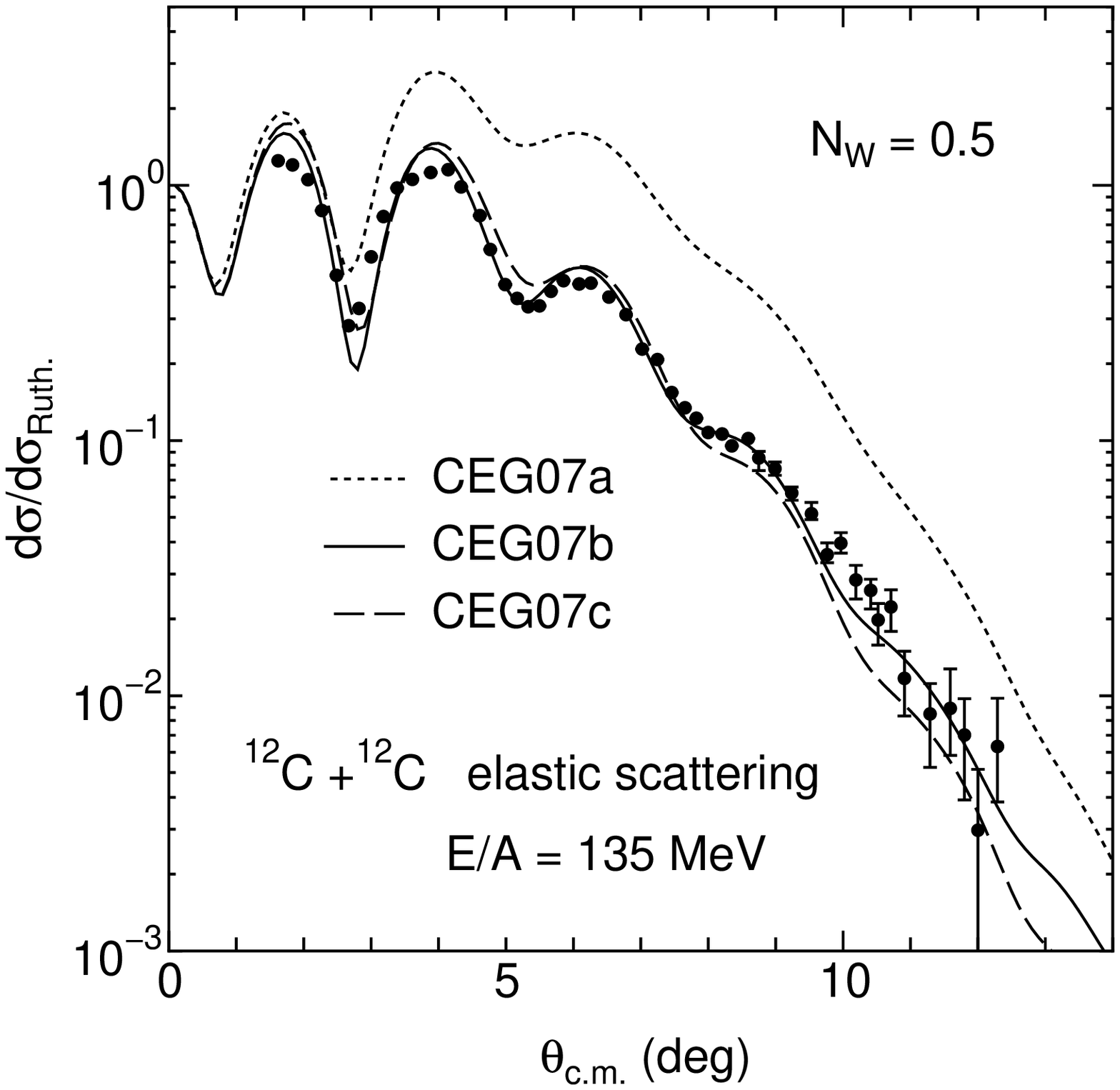}
\caption{
The cross section of the $^{16}$O +  $^{16}$O elastic scattering at $E/A$ = 70 MeV (left) and of the $^{12}$C + $^{12}$C one at $E/A$ = 135 MeV (right), respectively, displayed as the ratio to the respective Rutherford cross sections. 
The meaning of the curves are the same as that in Fig.~\ref{fig:CEG07POT}.
The cross sections are obtained with the DFM potentials in Fig.~\ref{fig:CEG07POT} but with the imaginary potential being multiplied by the renormalization factor $N_{\rm W}$ shown in the figures.
The figures are taken from Ref.~\cite{FUR09}.
}
\label{fig:CEG07XS}
\end{figure}

The new complex $G$-matrix is named as the CEG07~\cite{CEG07} ({\em Complex Effective potential with Gaussian form factors 2007}) that was derived from a modern NN interaction model in free space called the ESC04 ({\em Extended Soft-Core model 2004})~\cite{ESC1,ESC2}.
This is an improved version of the original CEG83 and CEG86 interactions proposed 30 years ago by Miyazaki group~\cite{CEG83,CEG86}.
The ESC04 is designed to give a consistent description of the interactions not only for the NN system but also for nucleon-hyperon (NY) and hyperon-hyperon (YY)
systems in the SU3 framework.
Therefore, the theoretical framework for CEG07 can easily extended to calculate the effective interactions for NY and YY systems in nucleon-hyperon mixed matter, such as those expected in neutron stars. 
It can also be applied to the study of hyperon-nucleus interaction both in the bound state (hyper nucleus) as well as in the scattering states~\cite{FUR10HY}. 

\subsection{Roles of three-body force in nucleus-nucleus scattering}

As is discussed above, the empirical saturation point cannot be reproduced by the lowest-order Brueckner theory (the $G$-matrix approximation), as long as one only uses the two-body NN interaction whatever the two-body NN interaction is.
This is also true in the case of the CEG07 $G$-matrix generated by the two-body force alone (which is called as CEG07a type).
This deficiency can be corrected clearly by introducing the three-nucleon force (3NF) composed of the three-body attraction (TBA) and the three-body repulsion
(TBR)~\cite{Lagaris81,Wiringa88,Baldo97}. 
The TBA is typically due to the two-pion exchange with excitation 
of an intermediate $\Delta $ resonance, that is the famous Fujita-Miyazawa diagram~\cite{FUJ57}.
An effective two-body interaction was derived from the TBA according to the formalism given in Ref.~\cite{Kasa74} and was added on the CEG07 $G$-matrix.
The attractive 3NF effect is known to give an important contribution at low densities.
However, the contribution of TBA alone does not resolve the problem of the saturation point. 

The role of the TBR is far more important than that of the TBA~\cite{FUR08}. 
The TBR contribution becomes more and more remarkable as the density becomes higher, which plays a decisive role for the shape of the saturation curve. 
It is well known that such a TBR effect is indispensable to obtain the stiff
equation-of-state (EOS) of the neutron-star matter assuring the observed
maximum mass of neutron stars. 
However, the origin of the TBR is not necessarily established. 
In the ESC approach, the TBR-like effects are represented rather phenomenologically
as the density-dependent two-body interactions induced by changing the vector-meson masses $M_{V}$ in nuclear matter according to 
$M_{V} (\rho) = M_{V} \exp({-\alpha_{V} \rho})$ with the parameter
$\alpha_{V}$. 
As mentioned in Ref.~\cite{ESC2}, this TBR-like contribution
introduced in ESC is found to be very similar to that of the
phenomenological TBR proposed in Ref.~\cite{Lagaris81} given as the density-dependent two-body repulsion.
The CEG07 interaction supplemented with the attractive (TBA) and repulsiv (TBR)
three-body force is called as the CEG07b and CEG07c interactions.
(The detailed difference between CEG07b and CEG07c was given in Ref.~\cite{FUR08,FUR09}.)

The CEG07 models were first successfully applied to the
analysis of proton-nucleus elastic scattering over a wide range
of incident energies and target nuclei~\cite{FUR08}
and later to the analyses of elastic scattering of nucleus-nucleus systems with great success~\cite{FUR09R,FUR09,FUR10,FUR12GP}.
It was also applied to the analyses of nucleus-nucleus inelastic scattering~\cite{TAKA10,HOR10,FUR13,QU15} with the microscopic coupled-channel formalism in which not only the diagonal potentials but also the coupling potentials that induce inelastic excitations of the colliding nuclei were obtained with the CEG07 $G$-matrix with the 3NF effects included.

\begin{figure}[t]
\centering\includegraphics[width=3.0in]{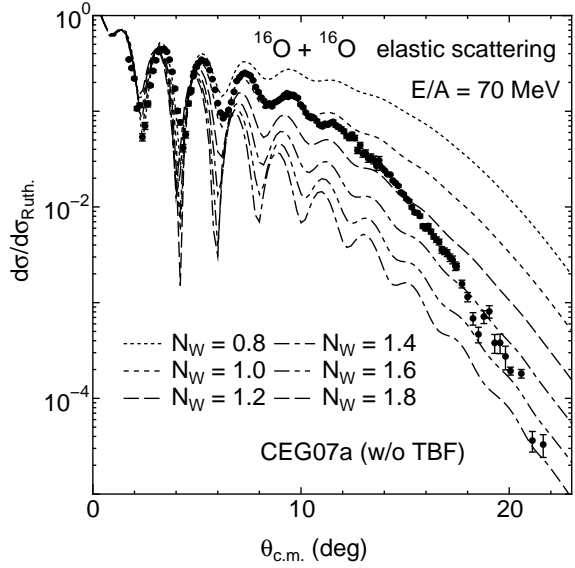}
\caption{
The cross sections for the $^{16}$O +  $^{16}$O
elastic scattering at $E/A$ = 70 MeV calculated by the DFM potetial obtained by CEG07a without the 3NF effect, by using various renormalization factors for the imaginary part.
The figure is taken from Ref.~\cite{FUR09}.
}
\label{fig:CEG07aXS}
\end{figure}

Figures~\ref{fig:CEG07POT} and \ref{fig:CEG07XS}  show typical examples that indicate the decisive role of the 3NF, particularly of its repulsive component TBR, on the calculated DFM potentials and elastic-scattering cross sections in the case of $^{16}$O +  $^{16}$O elastic scattering at $E/A$ = 70 MeV and  $^{12}$C + $^{12}$C one at $E/A$ = 135 MeV, respectively.
The dotted curves in Fig.~\ref{fig:CEG07POT} show the DFM potential calculated with the CEG07a $G$-matrix that contain no 3NF effect, while the dashed and solid curves are the results with CEG07b and CEG07c both inluding the 3NF effects. 
It is clearly seen that the 3NF strongly reduces the real potential strength in the spatial range of middle and short separations between the interacting nuclei where the local density defined by FDA (i.e. $\rho=\rho_{\rm P}+\rho_{\rm T}$) exceeds the saturation density $\rho_0$.
The 3NF effect on the imaginary part of the DFM potential is not so significant.

Figure~\ref{fig:CEG07XS} shows the elastic scattering cross sections calculated with the use of the DFM potentials shown in Fig.~\ref{fig:CEG07POT}.
The solid (CEG07b) and dashed (CEG07c) curves are the results with the 3NF effects that precisely reproduce the experimental data in whole angular range. 
The large deviation of the dotted curves (CEG07a), that include no 3NF effect, clearly indicates the important role of the 3NF effect.

It should be noted that, for each scattering system, the imaginary part of the DFM potential is multiplied by a common renormalization factor presented in the figures ($N_{\rm W}$ = 0.7 for $^{16}$O +  $^{16}$O system and $N_{\rm W}$ = 0.5 for $^{12}$C + $^{12}$C one) in calculating the cross section, where the $N_{\rm W}$ value is chosen so to optimize the fit of the solid curve to the experimental data.
The same $N_{\rm W}$ is used to all the three-kinds of calculations in each scattering system.
One might suspect that the large difference between the dotted curve and the other two curves could be compensated by using a more large $N_{\rm W}$ value.
In fact, the use of the larger $N_{\rm W}$ values drastically reduces the cross section at backward angles. 
However, it gives the angular slopes that completely different from those of the other two calculations with 3NF effects, as shown in Fig.~\ref{fig:CEG07aXS}~\cite{FUR09}. 
Thus, the large difference between the dotted curve and the other two curves really reflects the large difference of the real part of DFM potential that originates from the 3NF effects. 

These results clearly indicate that the elastic scattering of these light HI systems at intermediate energy is quite transparent enough to probe the difference of the depth and shape of the real potential at short distances, where the local density evaluated by the FDA can exceed the saturation value and the medium effect, particularly the 3NF effect, play an important role.  
In other words, it will be possible to fix some unknown parameter values relevant to the medium effects, such as the strength of 3NF, at high densities through the theoretical analyses of the experimental data of HI elastic scattering based on the present microscopic interaction model.

\section{Multi-Pomeron repulsion and the neutron-star mass}

The main subject of the present last section is to introduce the scenario recently proposed by Yamamoto {\em et al.}~\cite{MPa_EPJA16} who have connected the nucleus-nucleus scattering experiments in laboratories on the Earth and the maximum mass of neutron stars in the Universe obtained from the astronomical observations via the MPP interaction model.
We will briefly review their recent studies of investigating whether or not the maximum mass of $2M_\odot$ could be obtained from the EOS for hyperon-mixed neutron-star matter, when the effects of the three-body and four-body repulsion among the nucleons and hyperons are taken into account.

\vspace{5mm}
It has been made clear that elastic scattering of light HI systems such as the  $^{16}$O +  $^{16}$O and  $^{12}$C + $^{12}$C systems at intermediate energies around $E/A\approx 100$ MeV is very transparent and quite sensitive to the strength and shape of the real potential at short distances.
The DFM potential with the GEG07b with the 3NF effect well reproduces the experimental data of elastic-scattering cross sections.
The effect of 3NF, particularly its repulsive component, plays a crucial role in the proper evaluation of the interaction potential between finite nuclei at short distances where the local density exceeds the saturation value $\rho_0$ or reaches up to twice the saturation value $\rho\cong 2\rho_0$.
This implies that the elastic scattering can probe the medium effect of strong interaction in dense nuclear matter.
This may open up a new window of studying the properties of infinite nuclear matter as well as neutron-star matter through the theoretical analyses of nucleus-nucleus elastic scattering in laboratories on the basis of the Brueckner's $G$-matrix theory and the BHF framework.  

As mentioned in the introduction, the recent astronomical observations of the maximum mass of neutron star of twice the solar mass,  (1.97 $\pm$ 0.04)$M_\odot$~\cite{NS10} and (2.01 $\pm$ 0.04)$M_\odot$\cite{NS13}, imposes a severe condition for the stiffness of the equation of motion (EOS) of the neutron-star matter.
This suggests the existence of strong repulsive 3NF for sustaining such a large neutron-star mass, since it is known that the maximum mass evaluated by any BHF calculations with only the NN two-nucleon force (2NF) cannot exceed $\sim 1.5 M_\odot$.
On the other hand, the hyperon (Y) mixing in neutron-star matter, that is expected to be switched on around the neutron density of $\rho \approx (2\sim 3)\rho_0$, is known to bring about a remarkable softening of the EOS, which cancels the effect of repulsive 3NF effect for the maximum mass~\cite{NSM1,NSM2,NSM3,NSM4}.
One of the ideas to avoid this serious problem, called ``Hyperon puzzle in neutron stars'', is to consider that the 3NF-like repulsions {\em work universally} for YNN, YYN, YYY , as well as for NNN that was proposed by Nishizaki, Yamamoto and Takatsuka~\cite{NSM3}. 

The repulsive component of the 3NF adopted in the CEG07b and CEG07c type $G$-matrices was introduced rather phenomenologically in the form of density-dependent effective 2NF and its parameter was determined so that the saturation curve reproduced the empirical saturation point.
Such phenomenological model for the repulsive 3NF is not necessarily suitable for being extended to be used as the universal three-baryon force (3BF) among nucleons and hyperons.
As a possible candidate for such universal repulsion among multi-baryon systems, the {\em multi-pomeron exchange potential} (MPP)~\cite{MPP} was introduced as a model of universal repulsions among three and four baryons on the basis of the Extended Soft Core (ESC) baryon-baryon interaction model, that is the bare NN interaction model in free space used in the construction of the CEG07-type $G$-matrices.

\subsection{Pomeron-exchange model for many-body force}

The pomeron is a virtual particle that simulates the multi-gluon exchange among baryons and  thought of as being related to an even number of gluon-exchanges.
It is a kind of substantial particle closely related to the strong repulsive core in the baryon-baryon interactions.

Recently, the multi-pomeron-exchange potential (MPP)  model has been proposed by Rijken, Yamamoto and their collaborators~\cite{MPP,MPa_EPJA16,MPa1} that represents the repulsive component of the three-body and four-body potentials among three and four baryons including nucleons and hyperons.  
In the MPP model, the three- and four-body local potentials are derived
from the triple- and quartic-Pomeron vertexes. 
In order to be implemented into the BHF calculation for the EOS of baryonic matters,
the many-body forces are represented by the density ($\rho$) dependent two-body potential in a baryonic medium.
\begin{figure}[b]
\centering\includegraphics[width=2.5in,angle=0]{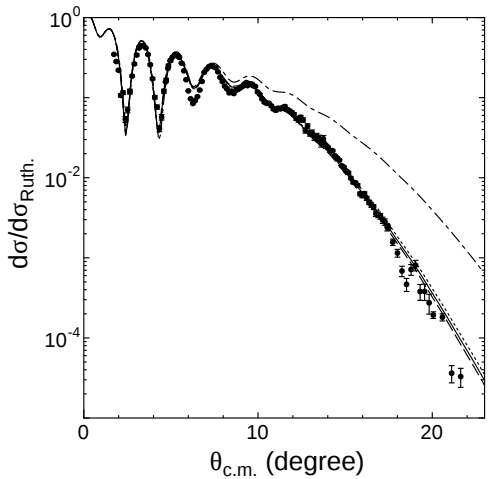}
\caption{Differential cross sections for $^{16}$O +  $^{16}$O elastic scattering
at $E/A$  = 70 MeV calculated with the $G$-matrix folding
potentials. Dot-dashed curve is the result with only the 2BF.
Solid, dashed and dotted curves are for MPa, MPa$^+$
and MPb, respectively. No renormalization in made on the imaginary part of the DFM potentials.
The figure is taken from Ref.~\cite{MPa_EPJA16}  }
\label{fig:MPPxs}
\end{figure}
It is obtained by integrating over coordinates of third (and
fourth) particles in the three-body (and four-body) potentials
as follows~\cite{MPa1}:
\begin{eqnarray}
V_{\rm eff}^{(N)} (\bm x_1, \bm x_2)\;  
 &=& \rho_{\rm NM}^{N-2} \int d\bm x_3  \cdot\cdot\cdot 
                                   \int d\bm x_N 
  V(\bm x_1, \bm x_2,  \cdot\cdot\cdot , \bm x_N) , 
\end{eqnarray}
where $V(\bm x_1, \bm x_2,  \cdot\cdot\cdot , \bm x_N)$ is the $N$-body local potential by Pomeron exchange.
In the case of $N=3$ (three-body) and $N=4$ (four-body), the effective $\rho$-dependent potentials have the following forms;
\begin{eqnarray}
V_{\rm eff}^{(3)} (r) &=& g_P^{(3)} \bigl( g_P^N \bigr)^3 \frac{\rho}{M^5} F(r) , \\ 
V_{\rm eff}^{(4)} (r) &=& g_P^{(4)} \bigl( g_P^N \bigr)^4 \frac{\rho^2}{M^8} F(r) , \\ 
F(r) &=& \frac{1}{4\pi} \frac{4}{\sqrt{\pi}} \Bigl( \frac{m_P}{\sqrt{2}} \Bigr)^3\:
\exp{ \Bigl( -\frac12 m_P^2 r^2   \Bigr)}\; .
\label{eq:MPP}
\end{eqnarray}
Here, $m_P$ and $ g_P^N$ are the pomeron mass and the two-body pomeron strength, that are taken to be same as those in ESC model for 2NF, and the scale mass $M$ is taken as a pomeron mass.

The MPP strengths $g_P^{(3)}$ and $g_P^{(4)}$ are unknown parameters that will be determined so that the experimental data are precisely reproduced with the DFM potential derived from the $G$-matrix including  MPP.
Namely, the $G$-matrix folding potentials including MPP contributions are
used to analyze the $^{16}$O +  $^{16}$O scattering at $E/A$ = 70 MeV and the strengths of MPP ($g_P^{(3)}$ and $g_P^{(4)}$) are adjusted so as to reproduce the experimental data.

\begin{figure}[b]
\centering
\includegraphics[width=4.2in,angle=0]{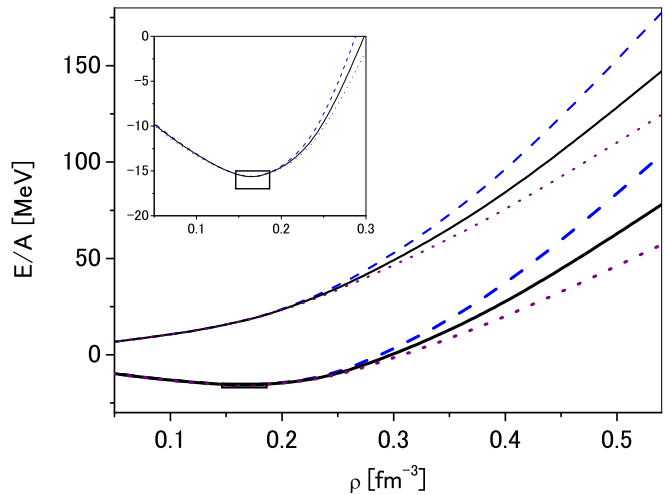}
\caption{Energy per particle ($E/A$) as a function of nucleon density $\rho$. 
The Upper (lower) curves are for neutron matter (symmetric nuclear
matter). The Solid, dashed and dotted curves are for MPa, MPa$^+$
and MPb, respectively. The box shows the empirical saturation point.
The inset shows a zoom of the region around the saturation point.
The figure is taken from Ref.~\cite{MPa_EPJA16}  
}
\label{fig:EOS}
\end{figure}

Figure~\ref{fig:MPPxs} shows the results.
In those calculations, the bare NN interaction ESC04 in free space has been replaced by the latest improved version called ESC08.
The dot-dashed curve are the result with the 2NF only that largely deviate from the experimental data, while other three curves include the many-body forces based on the MPP model.
The dotted curve (MPb) includes up to the 3NF, while the solid (MPa$^+$) and dashed (MPa) curves include both the 3NF and 4NF contributions. 
In the latter three curves, the MPP strengths $g_P^{(3)}$ and $g_P^{(4)}$ are determined so as to reproduce the data~\cite{MPa_EPJA16}. 
It should be noted that, in contrast with the DFM analyses with the CEG07 interactions, no renormalization is required to the imaginary part of the DFM potentials (i.e. $N_{\rm W}$ is fixed at 1.0) to attain the satisfactory fits to the data shown in the figure.

\subsection{Stiffness of Neutron-star matter fixed by nucleus-nucleus scattering}

The MPP interactions, the strength of which were determined from the analysis of nucleus-nucleus scattering, are then included in constructing the EOS of neutron-star matter as well as the symmetric nuclear matter and it is examined whether they result in an EOS stiff enough to give the observed large neutron-star mass~\cite{NS10,NS13}. 
It should be noted that the MPP is defined so as to work universally not only in NNN states but also YNN, YYN, and YYY states with the common parameters fixed by the analysis of nucleus-nucleus scattering.
Therefore, there is no additional free parameter for the interaction model. 
Corresponding to the determined MPP, the attractive component of 3NF (TNA) is added to reproduce the nuclear saturation property precisely.
Since TNA is important mainly in the lower density region, the addition of TNA does not lead to any additional ambiguity in the following discussion about the stiffness of hadronic matter in high density region.

Figure~\ref{fig:EOS} shows the obtained results for the saturation curves for the symmetric nuclear matter as well as for pure neutron matter~\cite{MPa_EPJA16}.
The upper curves are for the neutron matter and the lower curves are for the symmetric nuclear matter.
The solid, dashed and dotted curves are the result obtained with the MPa, MPa$^+$
and MPb interactions, respectively. 
The box shows the empirical value of saturation point. 
The inset shows a zoom of the region around the saturation point for the symmetric nuclear matter.
It is seen that the inclusion of the quartic-pomeron exchanges contributions (in MPa and MPa$^+$) in addition to the triple-pomeron exchange stiffen the saturation curve in the high density region in both the symmetric nuclear matter and neutron matter.
This is because the strengths of the effective two-body interaction derived from quartic-pomeron exchanges are proportional to $\rho^2$ (cf. Eq.~(\ref{eq:MPP})) and the contribution become sizable in the high-density region.

\begin{figure}[b]
\centering
{
\includegraphics[width=2.8in,angle=0]{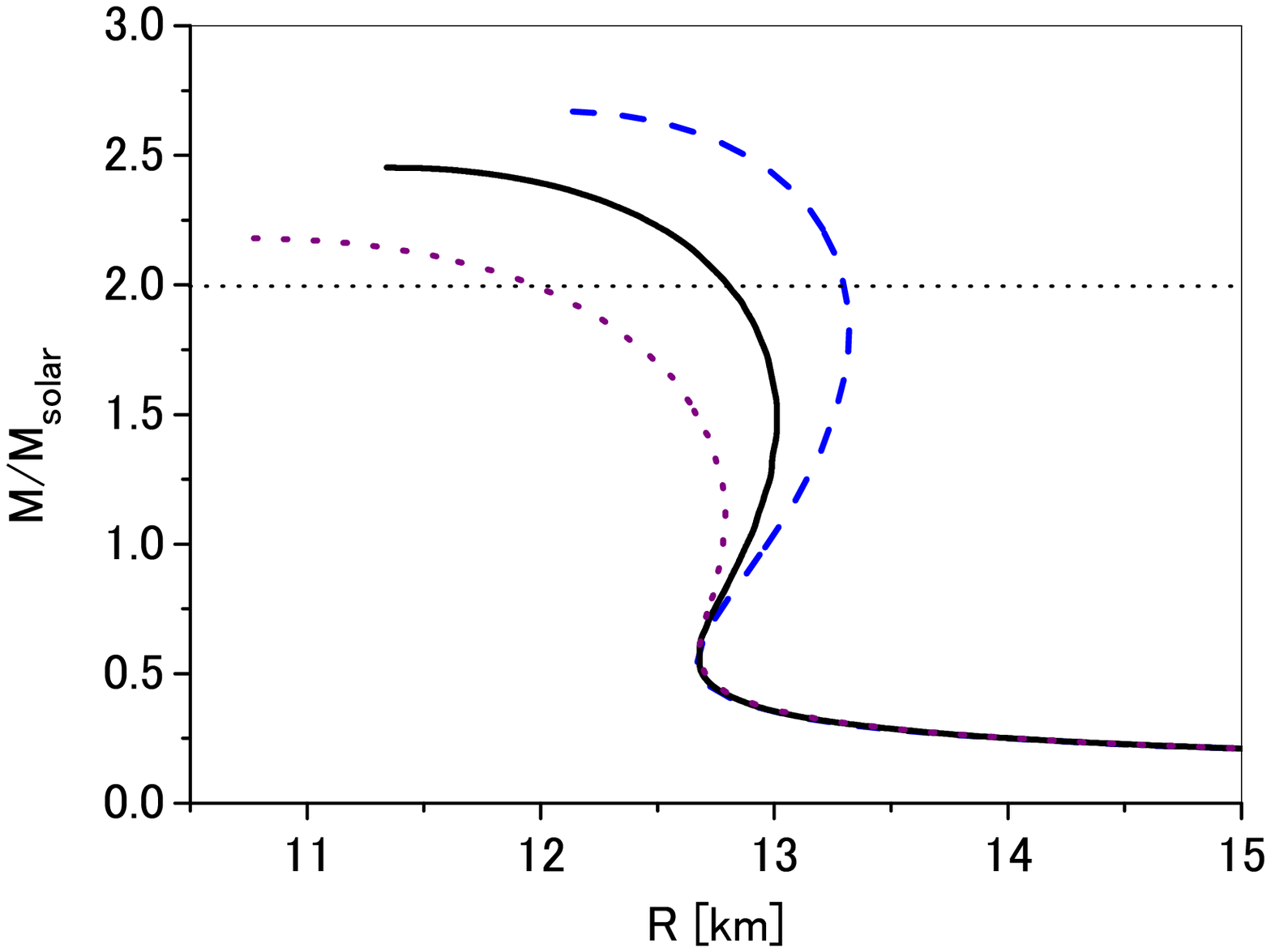}
\hspace{0mm}\includegraphics[width=2.8in,angle=0]{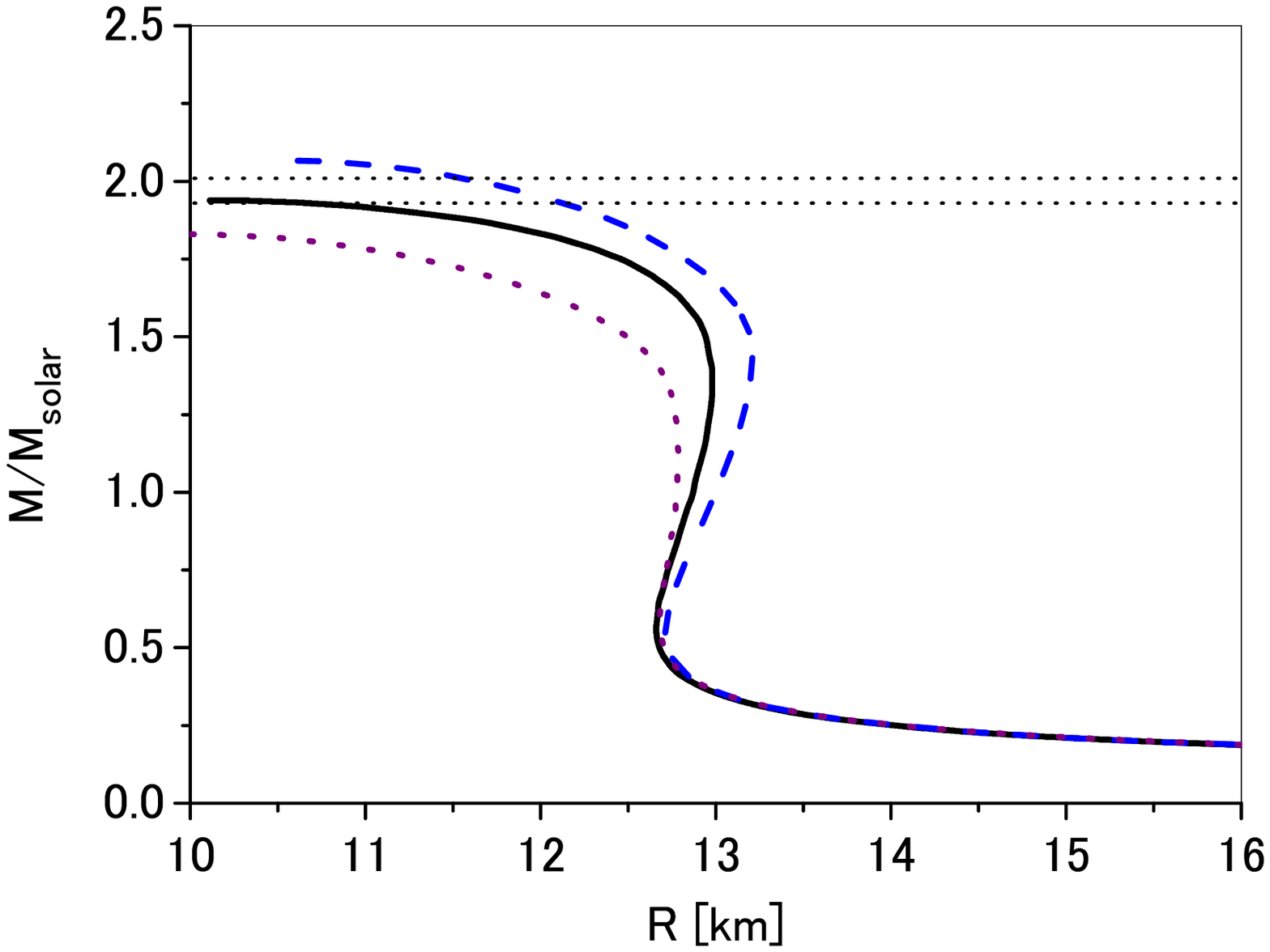}
}
\caption{Neutron-star masses as a function of the radius $R$.
Solid, dashed and dotted curves are for MPa, MPa$^+$ and MPb.
All the calculations in the right panel include the effect of hyperon mixing,
while those in the left panel are calculated for pure neutron matter without the hyperon mixing.
The  horizontal dotted lines show the observed mass $( 1.97 \pm 0.04 ) M_{\odot}$ of
J1614-2230.
The figures are taken from Ref.~\cite{MPa_EPJA16}  }
\label{fig:NMR}
\end{figure}

The final step is to study properties of neutron stars with and without hyperon mixing on the basis of these multi-baryon interaction model. 

Assuming a mixed matter of $n$, $p$, $e^-$ and $\mu^-$ in chemical equilibrium, the TOV equation is solved for the hydrostatic structure of a spherical non-rotating star~\cite{MPa_EPJA16}. 
The obtained mass-radius relations of neutron stars without hyperon mixing taken into account are demonstrated in the left panel of Fig.~\ref{fig:NMR} .
The solid, dashed and dotted curves are for MPa, MPa$^+$ and MPb,
respectively.
It is seen that the EOSs in these cases are found to be stiff enough to give $2M_{\odot}$.
The above differences appear significantly in the inner regions of $^{16}$O + $^{16}$O DFM potentials in the inner region, though they cannot be seen in the cross sections in Fig.~\ref{fig:MPPxs}.
It should be noted that the calculation with the 2NF alone leads to too soft EOS to give $2M_{\odot}$, though not shown here.

However, the hyperon mixing will be set on when the neutron density becomes as high as $(2\sim 3)\times M_{\odot}$, that will be expected to soften the EOS substantially. 
The EOS of $\beta$-stable neutron-star matter composed of neutrons ($n$), protons ($p^+$), electrons ($e^-$), muons ($\mu^-$), and hyperons ($\Lambda$ and $\Sigma^-$) is derived from the G-matrix calculation with the present interaction model. 
Using the EOS of hyperonic neuron-star matter, the Tolmann-Oppenheimer-Volkoff (TOV) equation for the hydrostatic structure is solved and the mass-radius relations of neutron stars is obtained~\cite{MPa_EPJA16}.
The results are shown in the right panel of  Fig.~\ref{fig:NMR}.

Calculated values of maximum masses for MPa$^+$, MPa and MPb are 2.07$M_{\odot}$, 1.94$M_{\odot}$ and 1.83$M_{\odot}$, respectively, being smaller by 0.61$M_{\odot}$, 0.51$M_{\odot}$ and 0.35$M_{\odot}$, than the values without hyperon mixing. 
Although the universal repulsion works to raise the maximum mass, the hyperon mixing also is enhanced by it so that the maximum mass is reduced. 

The maximum mass for MPb is considerably smaller than the observed value 
of $\sim 2M_{\odot}$. 
On the other hand, those for MPa and MPa+ reach to this value owing
to the four-body repulsive contributions.

\section{Concluding remark}

Nuclear physics provide an unique play ground to survey the properties of dense hadronic matter including those in neutron stars in the Universe. 
Precise measurements of refractive nucleus-nucleus scattering in laboratories on the Earth give us a quite valuable information about the high-density nuclear matter and provide a useful test ground of the effective interactions in high-density baryonic matter that act not only between two nucleons (NN) but also those among triple and quartic baryons (BBB).
The key bridge to connect the nuclear experiment in laboratories and the baryonic matter in the universe is the modern theories of the effective interaction in  many fermion systems. 
The new Brueckner's $G$-matrix constructed with the latest models for bare NN interaction in free space together with the recently developed many-body forces effective in high-density baryonic matter is a promising candidate for the bridge.  

\vspace{5mm}
I should like to dedicate the present review article to the memory of late Professor Yoichiro Nambu.

\section*{Acknowledgment}

The present review article is written along a series of works with Takenori Furumoto, Yasuo Yamamoto, Nobutoshi Yasutake, Thomas A.~Rijken, Masahiko Katsuma, Shigeto Okabe, Yoshio Kondo, Masaaki Takashina, Wataru Horiuchi as well as experimental group at Beihang University/RCNP Osaka University headed by Isao Tanihata. I should like to thank all the collaborators for their contributions.
A part of the present work was supported by JSPS KAKENHI Grant Number 15K05087.





%

\end{document}